\documentclass[a4paper,11pt]{article}
\pdfoutput=1  

\usepackage{jheppub}  

\usepackage[T1]{fontenc} 

\usepackage{lipsum}  

\usepackage{accents}

\usepackage{comment}

\usepackage{enumitem}

\newcommand{\be}{\begin{equation}}
\newcommand{\ee}{\end{equation}}

\newcommand{\nn}{\nonumber}

\newcommand{\cF}{\mathcal{F}}
\newcommand{\cG}{\mathcal{G}}
\newcommand{\cH}{\mathcal{H}}
\newcommand{\cI}{\mathcal{I}}

\newcommand{\cK}{\mathcal{K}}

\newcommand{\cM}{\mathcal{M}}

\newcommand{\cO}{\mathcal{O}}

\newcommand{\cR}{\mathcal{R}}
\newcommand{\cS}{\mathcal{S}}
\newcommand{\cT}{\mathcal{T}}

\newcommand{\cV}{\mathcal{V}}

\title{\boldmath Consistent truncations from the geometry of sphere bundles}

\author[a]{Federico Bonetti,}
\author[b]{Ruben Minasian,}
\author[c]{Valent\'i Vall Camell,}
\author[d]{and Peter Weck}

\affiliation[a]{Department of Mathematical Sciences, Durham University, Durham, DH1 3LE, United Kingdom}
\affiliation[b]{Institut de Physique Th\'{e}orique, Universit\'{e} Paris Saclay, CNRS, CEA, F-91191, Gif-sur-Yvette, France}
\affiliation[c]{Arnold Sommerfeld Center for Theoretical Physics, Ludwig-Maximilians Universit\"{a}t, Theresienstra{\ss}e 37, 80333 M\"{u}nchen, Germany}
\affiliation[d]{Department of Physics and Astronomy, Johns Hopkins University, 3400 North Charles Street, Baltimore, MD 21218, USA}

\emailAdd{federico.bonetti@durham.ac.uk, ruben.minasian@ipht.fr, valenti.vallcamell@gmail.com, pweck1@jhu.edu}

\abstract{In this paper, we present a unified perspective on sphere consistent truncations based on the classical geometric properties of sphere bundles. 
The backbone of our approach is the global angular form for the sphere. 
A universal formula for the Kaluza-Klein ansatz of the flux threading the $n$-sphere captures the full nonabelian isometry group $SO(n+1)$ and scalar deformations associated to the coset $SL(n+1,\mathbb R)/SO(n+1)$.
In all cases, the scalars enter the ansatz in a shift by an exact form. We find that the latter can be completely fixed by imposing mild conditions, motivated by supersymmetry, on the scalar potential arising from dimensional reduction of the higher dimensional theory.
We comment on the role of the global
angular form in the derivation
of the topological couplings
of the lower-dimensional theory,
and on how this perspective could provide
inroads into the study of consistent
truncations with less supersymmetry.
}

\begin{document} 
\begin{flushright}
LMU-ASC 54/22
\end{flushright}
\maketitle
\flushbottom

\section{Introduction}
\label{sec_intro}
Consistency of truncations, \emph{i.e.}~ensuring that any solution of a lower-dimensional theory obtained via Kaluza-Klein (KK) reduction automatically satisfies the higher-dimensional equations of motion, is of obvious practical interest. The spectrum of the lower-dimensional theory is determined by linearized KK analysis. However, nonlinear modifications away from the infinitesimal neighbourhood of the ground state are required in order to capture the interactions in the reduced theory and ensure that all higher-dimensional equations of motion are satisfied. The problem has received much recent (and not so-recent) attention, and many instances of consistent truncation have been elaborated in the literature. The conditions for a given truncation to be consistent are relatively straightforward when the reduction is on a group manifold, or a quotient theoreof: only the modes invariant under the group action should be kept \cite{Scherk:1979zr}. However, the conditions for consistent truncation for more general internal manifolds are significantly less well understood. 

This paper is an attempt to present a unified and synthetic approach to the construction of consistent truncations, based on the geometric properties of the fibre bundle employed in the reduction.  We shall discuss here only the cases of sphere reductions (collected in Table~\ref{table}), where our formulae can be compared with known uplift results \cite{Pilch:1984xy,Nastase:1999cb,Nastase:1999kf,deWit:1983vq,deWit:1984nz,deWit:1985iy,deWit:1986mz,deWit:1986oxb,Nicolai:2011cy,deWit:2013ija,Hohm:2013pua,Godazgar:2013dma,Hohm:2014qga,Godazgar:2015qia,Varela:2015ywx,Kim:1985ez,Cvetic:1999xp,Lu:1999bw,Cvetic:1999xx,Cvetic:2000eb,Khavaev:1998fb,Cvetic:2000nc,Pilch:2000ue,Cassani:2010uw,Liu:2010sa,Gauntlett:2010vu,Skenderis:2010vz,Baguet:2015sma,Guarino:2015jca,Guarino:2015qaa, Guarino:2015vca,Cvetic:2000dm}.
Our analysis retains the gauge fields for the full nonabelian isometry group of the sphere as well as scalar fields encoding its deformation. The justification and content of our approach can be summarized in three main points:

\begin{enumerate}[left=- 0.7\parindent .. 0pt, listparindent=\parindent, parsep=0pt]
\item 
The popular approach, mostly based on variants of exceptional field theory 
(see \emph{e.g.}~\cite{Lee:2014mla,Hohm:2014qga,Baguet:2015sma,
Ciceri:2016dmd,Cassani:2016ncu,Berman:2020tqn,Malek:2020yue}), solves the problem of combining the lower-dimensional fields into their higher-dimensional progenitors using clever applications of representations of duality symmetries.
In contrast, the starting point of our analysis is a classical geometric object---the global angular form \cite{bott1982differential} $e_n$ on an $n$-sphere bundle over the $AdS$ space in question. 
The naive space of deformations on an $n$-sphere is given by $SL(n+1,\mathbb R)/SO(n+1)$, and the only group-theoretic fact required is that the denominator and the numerator of this coset embed maximally into the duality group and its maximal compact subgroup, respectively.\footnote{In case of IIB reductions, one needs to incorporate $SL(2)$ and $U(1)$ factors, respectively. However, they can and will be ignored here.} In other words, our approach offers an alternative and in some ways simpler starting point for constructing a given consistent truncation ansatz.

When restricted to each fibre, $e_n$ is the generator of the top cohomology of the fibre, \emph{i.e.}~the volume form. The pull-back of its exterior derivative gives the Euler class of the sphere bundle (notably it vanishes for even $n$). More details can be found in section \ref{GAF_review}. Crucially, for any given sphere consistent truncation, the flux threading the internal sphere can be reconstructed from this object---or rather, as we shall see, from an appropriate incarnation of it, which we denote $e_n'$.

\item 
The embedding of $e_n'$ into the consistent truncation ansatz (see section \ref{sec_uplift}) allows us to see several important universal features. 

First, the flux through the undeformed $n$-sphere bundle is always of the form
\be \label{eq:start}
\hat \cF_n |_{\mbox{\tiny no defs}}    =  e_n'  .
\ee
This can be thought of as a common starting point for all the sphere consistent truncations.  Away from the vacuum solution, the sphere is deformed and the solution needs to accommodate the associated scalar modes, which are encoded in a symmetric unimodular matrix, $T$. The inclusion of these scalars simply amounts to shifting $e_n'$ by an exact form. On a case-by-case basis there can also be contributions from extra fields. 

While an exact shift to $e_n'$ would in principle still allow the addition of an infinite number of new terms, we demonstrate that the ambiguity can be fixed by considering the contributions of the kinetic terms for $\hat \cF_n $ to the scalar potential.  General supersymmetry arguments tell us that the scalar potential is quadratic in $T$ (see section \ref{sec_integral} for details). 
The part of  the integral of $|\hat \cF_n|^2$ on the $n$-sphere that yields no-derivative terms is not manageable, and in general is not polynomial in $T$. However this is not the only contribution to the scalar potential. The other comes from the reduction of the higher dimensional Hilbert-Einstein term. This contribution is even less pleasant, but it admits a choice of parameters from the deformed higher-dimensional metric that collects all non-polynomial (non-quadratic) contributions into the integral of a perfect square, while still leaving a free parameter. This contribution can be cancelled against the contribution from the flux kinetic term, provided the final free parameter is set to zero. Remarkably, by choosing the values for these free parameters as described above, we land precisely on the exact shift to $e_n'$ and on the values for $n$ and $D$ (the dimension $D$ of the full theory) which are required for consistent truncation.

As an aside, these statements can be compared with those in \cite{Lee:2014mla} based on the generalised parallelizability of spheres. There trivial $SO(n+1)$ connections were considered. In that case, $e_n' = {\rm vol}_n$, and the shift to \eqref{eq:start} is indeed exact, while the  relevant parameters are fixed based on the requirement of off-shell supersymmetry, to the same values found here.
Note that arguments based
on the scalar potential
of the lower-dimensional
theory have also appeared 
in \cite{Nastase:2000tu}.

\item 
Equation \eqref{eq:start} is already sufficient for obtaining lower-dimensional topological Chern-Simons couplings.\footnote{This is true \emph{e.g.}~for $AdS_7$ and $AdS_5$ reductions,
as in \cite{Freed:1998tg,Harvey:1998bx}.
In $AdS_4$ the Chern-Simons couplings involve scalar fields, and require extra care.} Hence one way of describing our approach would be by saying that it provides a consistent completion of the reduction in the topological sector. 
The latter in turn functions as a backbone
for the full consistent Kaluza-Klein ansatz.
This perspective could offer a fruitful
counterpart to approaches rooted
in exceptional duality groups,
whose details are in general
dependent on the dimensions
and the amount of supersymmetry.

For many $AdS_5$ theories with less supersymmetry, where the lower-dimensional symmetry based approaches are less powerful, the Chern-Simons couplings can and have been obtained by suitable modifications of  $e_n'$
\cite{Bah:2019rgq,Bah:2020jas,Bah:2020uev}. In other words, in all these cases a new starting point similar to \eqref{eq:start} is already available. While there are a number of technical challenges in completing to the full solution from this starting point, better understanding the general construction of $e_n'$ with scalar deformations and its embedding into the consistent truncation ansatz are worth exploring.

\end{enumerate}

\section{Global angular forms and sphere truncations}
\label{sec_uplift}

\begin{table}
    \renewcommand{\arraystretch}{1.5}
    \small
    \centering
    \begin{tabular}{ l | c | c | c | c |  c c }
         \multicolumn{1}{c|}{ Higher dimensional theory }   &  Sphere & Flux $\hat \cF$   & Dilaton & Gauge group  & Extra fields \\ \hline \hline
       \parbox[c]{3.8cm}{
      bosonic sector of \\
      $D=11$ sugra
       }  \rule[-4mm]{0mm}{10mm} &  $S^4$ & $\hat G_4$ & no & $SO(5)$ 
         & 5 three-forms\\
         \hline
        \parbox[c]{3.8cm}  {
bosonic sector of\\
$D=11$ sugra
} \rule[-4mm]{0mm}{10mm}  &  $S^7$  & $\hat * \hat G_4$ & no & $SO(8)$   &  35 pseudoscalars \\
\hline 
        \parbox[c]{3.8cm}{
$SL(2,\mathbb R)$ singlet sector of\\
$D=10$ type IIB sugra 
}  \rule[-4mm]{0mm}{10mm}  &  $S^5$ &   $\hat F_5$ & no & $SO(6)$  &  none \\
        \hline
     \parbox[c]{3.8cm}  {
bosonic  sector of massive\\
$D=10$   type IIA sugra
}  \rule[-4mm]{0mm}{10mm}  &  $S^6$
        & $\hat * \hat F_4$ & yes & $ISO(7)$ & see sec.~\ref{sec_massive} \\
\hline 
        \eqref{starting_action} with $p=2$ in $D$ dim. 
        \rule[-4mm]{0mm}{10mm}
        & $S^2$  & $\hat F_2$ & yes & $SO(3) $ & none \\ 
        \hline
        \eqref{starting_action} with $p=3$ in $D$ dim. 
         \rule[-4mm]{0mm}{10mm} & $S^3$  & $\hat F_3$ & yes & $SO(4)$ & 1 two-form\\
      \hline  
        \eqref{starting_action} with $p=3$ in $D$ dim.
         \rule[-4mm]{0mm}{10mm} & $S^{D-3}$  & $ \hat * F_3$  & yes & $SO(D-2)$ & none   
    \end{tabular}
    \caption{Consistent truncations on spheres considered in this section. 
For each case, the higher-dimensional theory is
indicated, as well as the dimensionality of the sphere used in the truncation. The quantity $\hat \cF$ denotes the $D$-dimensional flux that
threads the sphere; in each case we
indicate which field strength
(or Hodge dual thereof) in the $D$-dimensional theory is identified with $\hat \cF$.  The column `Dilaton' indicates whether the kinetic term for $\hat \cF$ in the
$D$-dimensional theory comes with a dilaton prefactor. We also indicate the gauge group
of the lower-dimensional gauged model.
Finally, the column `Extra fields' collects the bosonic fields that have to be added 
for consistency, in addition to the gauge fields
of $SO(n+1)$ and the scalars parametrizing
the coset $SL(n+1, \mathbb R)/SO(n+1)$.
}
    \label{table}
\end{table}

In this section we consider
the sphere consistent truncations
listed in Table \ref{table}.
Notice that for 11d supergravity on $S^4$ and $S^7$,
for type IIB supergravity on $S^5$, and for
massive type IIA supergravity on $S^6$
the consistent truncation applies to the full
content of the model, including fermions.
For simplicity, throughout this work we 
restrict to the bosonic sectors
(and, for type IIB, to the $SL(2,\mathbb R)$
singlet sector consisting of the Einstein frame metric and the self-dual five-form flux).
We use $D$ to denote the spacetime dimension of the higher-dimensional model,
$d$ for the dimension of the lower-dimensional,
and $n$ for the dimension of the sphere
(so that clearly $D = d+n$).
We use a hat for  fields in $D$ dimensions.
The $D$-dimensional theory
contains, among other fields, an 
$n$-form field strength that 
threads the $S^n$, which we denote $\hat \cF_n$.

For the fourth, fifth, and sixth entry
in Table \ref{table} the starting $D$-dimensional theory is a   bosonic model
with action
\be \label{starting_action}
S_{(D)} = \int \bigg[ 
\hat R \hat * 1 
- \frac 12 d \hat \phi \wedge \hat * d \hat \phi
- \frac 12 e^{-a \hat \phi} \hat F_p \wedge \hat * \hat F_p 
\bigg ] \ , 
\ee 
where $\hat R \hat * 1$ is the standard
Einstein-Hilbert term,
$\hat \phi$ is a real scalar (dilaton),
and $\hat F_p$ is the closed field strength
of a $(p-1)$-form gauge potential. 
The positive constant $a$ is given by
\be   \label{a_def}
a^2 = 4 - \frac{2(p-1)(D-p-1)}{D-2}
= 4 - \frac{2(n-1)(D-n-1)}{D-2}\ ,
\ee 
which is the value required 
for the consistent truncation to be possible
\cite{Cvetic:2000dm}.
In the second step, we have observed that
the dimensionality $n$ of the sphere
equals $p$ if the flux threading the sphere
is $\hat F_p$, and $D-p$ if it is its Hodge
dual $\hat * \hat F_p$. In both cases
we get the same value of 
$a^2$ as a function of $D$, $n$.

In each case, the modes retained
in the truncation include the following
bosonic fields in the $d$-dimensional
supergravity theory:
\begin{itemize}
\item metric;
\item gauge fields associated to the $SO(n+1)$ isometry of the round sphere $S^n$;
\item real scalars parametrizing the coset space $SL(n+1,\mathbb R)/SO(n+1)$;
\item a real dilaton, if there is a `yes' 
in the pertinent column of Table \ref{table}.
\end{itemize}
Crucially,  consistency of the truncation
might require that we keep 
additional modes in $d$ dimensions,
as reported in the last column of
Table \ref{table}.

Our main objects of interest
are the uplift formulae for the $D$-dimensional
metric, dilaton, and $n$-form flux $\hat \cF_n$
 that threads the sphere.
In particular, we focus on the parts
of the uplift formulae that only contain
the $SO(n+1)$ gauge fields
and the $SL(n+1,\mathbb R)/SO(n+1)$ scalars.
We observe a simple regular pattern,
described in detail below,
which extends the analysis of
\cite{Nastase:2000tu, Lee:2014mla}
by considering simultaneously the
$SO(n+1)$ gauge fields and the $SL(n+1,\mathbb R)/SO(n+1)$ scalars.

The detailed forms of the uplift formulae
depend on the specific case considered;
we describe each case in turn in the following subsections. 
The general structure of the uplift formulae, however, can be presented
uniformly for all cases.
The $D$-dimensional metric, dilaton, and $n$-form flux $\hat \cF_n$
take the form
\begin{align}
d\hat s^2_D  & =Y^{c_1} (yTy)^{\frac{n-1}{D-2}}  \bigg[  ds^2_d 
+   g^{-2} Y^{- \frac{2}{n+1}} \frac{1}{yTy} T^{-1}_{IJ} Dy^I Dy^J \bigg]  
+ \text{(extra)}  \ ,
\label{general_metric}\\
e^{\frac{2s}{a} \hat \phi} & = Y^{c_2} (yTy)^{-1}
\times \text{(extra)} \  , \label{general_dilaton}\\
\hat \cF_n   & =  e_n' + \frac{1}{\cV_n n!} d \bigg[
\frac{n}{n-1} \frac{1}{yTy}
\epsilon y (Ty) (Dy)^{n-1}
\bigg] + \text{(extra)}   \ , \label{general_flux}
\\
e'_n & =\frac{1}{\cV_n n!} \epsilon y (Dy)^n
+ \frac{1}{\cV_n n!}
\sum_{j=1}^{j_{\rm max} }  \frac{g^j}{j!} 
(n/2)_j
\epsilon y F^j (Dy)^{n-2j} \ .
\label{general_e_prime}
\end{align}
Our notation is as follows.
We use $ds^2_d$ for the metric
in $d$ dimensions,
while $Y$ is a positive scalar
that encodes the $d$-dimensional dilaton, if present
(if absent, 
the above formulae are understood with $Y \equiv 1$).
The 
positive constant $a$ was defined in 
\eqref{a_def} while the
constants $c_1$, $c_2$ are given by 
\be  \label{c1c2def}
c_1 = \frac{2n}{(D-2)(n+1)} \ , \qquad
c_2 = \frac{1}{a^2}  \frac{4(D-n-2)(n-1)}{(D-2)(n+1)} \ .
\ee 
The quantity $s$ is a sign, determined
by how the $D$-dimensional dilaton enters
the kinetic term for the flux that threads the sphere,
$S_{(D)} \sim \int e^{-s a \hat \phi} \hat \cF_{n} \wedge \hat * \hat \cF_n + \dots$.
The symbol $\cV_n$ stands for the volume
of the  round sphere $S^n$ of radius 1,
\be  \label{unit_sphere}
\cV_n = \frac{2 \pi^{\frac{n+1}{2}}}{\Gamma(\frac{n+1}{2})}  \ ,
\ee 
while  $(n/2)_j$ is the descending Pochhammer symbol, given by $(n/2)_0 :=1$ and
\be   \label{pochhammer_def}
(n/2)_j = \frac n2 \left( \frac n2 -1 \right)
\left( \frac n2 -2 \right)\dots \left(\frac n2 - j +1 \right) \ .
\ee

The indices $I$, $J$, $K = 1,\dots,n+1$ are vector indices
of $SO(n+1)$. Unless otherwise stated, they are raised and lowered with the $SO(n+1)$ invariant tensor $\delta_{IJ}$
and its inverse $\delta^{IJ}$.
The quantities $y^I$ are 
constrained coordinates on the sphere $S^n$.
They are defined via the standard
embedding of the round unit $S^n$
in flat $\mathbb R^{n+1}$ in
Cartesian coordinates, and therefore satisfy
\be \label{sphere_eq}
\delta_{IJ} y^I y^J = 1 \ .
\ee
The 1-forms $Dy^I$ are defined as
\be \label{Dy_def}
Dy^I = dy^I + g A^{IJ} y_J \ ,
\ee
where the 1-forms $A^{IJ}  =A^{[IJ]}  $ are the 
external $d$-dimensional
gauge fields of $SO(n+1)$,
and $g$ is a constant parameter.
The associated field strength reads
\be 
F^{IJ} = dA^{IJ} + g A^{IK} \wedge A_K{}^J \ .
\ee
The quantities $T_{IJ}$ 
are the entries
of a symmetric, positive definite, unimodular matrix,
\be 
T_{IJ} = T_{JI} \ , \qquad \det T_{IJ} =1 \ .
\ee 
The matrix $T$ depends only on the external $d$-dimensional spacetime coordinates and 
parametrizes the $n(n+3)/2$ real scalars of the coset
$SL(n+1,\mathbb R)/SO(n+1)$.

In 
\eqref{general_metric}-\eqref{general_e_prime}
we have introduced some shorthand notation
used throughout this work. First of all,
we write
\be  \label{index_contractions}
yTy = y_I T^{IJ} y_J \ , \qquad (Ty)^I = T^{IJ} y_J \ .
\ee 
Secondly, we have suppressed $I$ indices
and wedge products of differential forms
in expressions involving the epsilon tensor
of $SO(n+1)$. For example,
\begin{align}  \label{shorthand}
\epsilon y (Ty) (Dy)^{n-1} & = \epsilon_{IJK_1 \dots K_{n-1}} y^I (Ty)^J Dy^{K_1} \wedge \dots \wedge Dy^{K_{n-1}} \ , \\
\epsilon y F^2 (Dy)^{n-4}
& = \epsilon_{L J_1 K_1 J_2 K_2 I_1 \dots I_{n-4}} y^{L} F^{J_1 K_1} \wedge F^{J_2 K_2}
\wedge Dy^{I_1} \wedge \dots \wedge Dy^{I_{n-4}} \ . 
\end{align}

As mentioned above, the quantity
$\hat \cF_n$ is the flux threading the $S^n$.
We normalize $\hat \cF_n$ so that 
its flux through $S^n$ is one,
\be 
\int_{S^n} \hat \cF_n = 1 \ .
\ee 
The physical flux is in general an integer
multiple of $\hat \cF_n$.
We comment on the structure of $e_n'$ in greater detail below.
Interestingly, $e'_{n}$
does not contain the external scalars $T_{IJ}$. The latter enter $\hat \cF_n$ only
via the total derivative term in \eqref{general_flux}.
This is an essential feature 
of the presentation 
\eqref{general_flux} of the uplift
formula for the $n$-form flux.
It has been derived in \cite{Lee:2014mla}
in the case in which the gauge
fields of $SO(n+1)$ are set to zero.\footnote{
The relative factor between
$\epsilon y(Dy)^n$ and
$\frac{1}{yTy} \epsilon y(Ty)(Dy)^{n-1}$
in \eqref{general_flux}, \eqref{general_e_prime} is $\frac{n}{n-1}$,
correcting a typo in equation (2.50)
of \cite{Lee:2014mla} (the factor $2(d-2)!$
in the denominator of $A'$  
should be $(d-1)!$).
}
Furthermore, we note that the $d$-dimensional dilaton $Y$
does not enter the uplift formula
for $\hat \cF_n$.

As pointed out above,  retaining
only the metric, the gauge fields $A^{IJ}$,  the scalars
$T^{IJ}$, and the dilaton $Y$
(if there is a `yes' entry in Table \ref{table}) 
in the lower-dimensional model
might not be consistent, and extra fields might be needed (see last column of Table \ref{table}). When extra fields are present,
they generically
enter the uplift formulae
for $d \hat s^2_D$, $\hat \phi$, and/or
$\hat \cF_n$.
We have used the notation 
$ \text{(extra)}$ in \eqref{general_metric}-\eqref{general_flux}
as a reminder of this caveat.

The quantity $e_n'$ is closely related to
the canonical global angular form
for an $S^n$ bundle.
Therefore, before proceeding, we briefly
review the salient aspects of global angular forms.

\subsection{Brief review of global angular forms}

\label{GAF_review}

We follow \cite{Harvey:1998bx}, see also
the textbook \cite{bott1982differential}.
Let $E$ be a real oriented vector bundle of
rank $n+1$ over a base manifold $B$.
Suppose $E$ is equipped with a connection and
a metric.
Let $S(E)$ denote the associated
unit-sphere bundle: if $p\in B$
and we choose Cartesian coordinates
$y^I$, $I = 1,\dots,n+1$ on the $\mathbb R^{n+1}$
fiber of $E$ at $p$, the sphere fiber of $S(E)$ at $p$
is described by \eqref{sphere_eq}.
We use $E_0$ to denote the complement
of the zero-section in $E$.
One can prove that
there exists a globally defined $n$-form 
on $E_0$, 
which we denote $e_n$\footnote{In the mathematics literature the global angular form is usually normalized in such a way that it integrates to 2
on the sphere fibers, so that $e_n^{\rm here} = \tfrac 12 e_n^{\rm maths}$. For $n$ even, 
the closed form $e_n^{\rm maths}$ represents
an \emph{integral} cohomology class.
},
with the following properties:
\begin{itemize}
\item The form $e_n$ restricted to the fibers of 
$S(E)$ reduces to the standard volume form on $S^n$,
normalized to integrate to $1$,
\be 
\int_{S^n} e_n = 1 \ .
\ee 
\item The exterior derivative of $e_n$ is given by
\be 
de_n = \left\{  
\begin{array}{ll}
0 & \text{if $n$ is even,} \\
-   \pi^* \chi_{n+1}(E) \qquad  & \text{if $n$ is odd.}
\end{array}
\right.
\ee  
Here 
$\pi^*$ is the pullback by the projection
$\pi : E_0 \rightarrow B$ onto the base space. The $(n+1)$-form    
$\chi_{n+1}(E)$ is a polynomial
in the curvature of the bundle $E$, given
explicitly below in \eqref{chi_eq}, that represents
the Euler characteristic of $E$
as  a cohomology class in $H^{n+1}(B;\mathbb Z)$.
\end{itemize}
We can exhibit  an explicit local expression for $e_n$
in terms of the constrained coordinates
$y^I$ and the components $F^{IJ}$ of the field strength of the $SO(n+1)$
connection on $E$ \cite{Harvey:1998bx},
\be  \label{compact_en}
e_{n} = \frac{1}{\cV_n n!}
\sum_{j=0}^{\lfloor n/2 \rfloor }  \frac{g^j}{j!} 
(n/2)_j
\epsilon y F^j (Dy)^{n-2j} \ .
\ee  
The unit sphere volume $\cV_n$
and the Pochhammer symbols
$(n/2)_j$ have been
introduced in \eqref{unit_sphere},
\eqref{pochhammer_def}.

The term with $j=0$ in
\eqref{compact_en}
of schematic form
$\epsilon y (Dy)^n$
describes
 the volume form
on the round $S^n$,
normalized to integrate to 1.
The derivative of the $j=0$ term generates
a term with one $F$ factor,
schematically
$\epsilon y F (Dy)^{n-1}$.
The relative coefficient
between the $j=0$
and $j=1$ term in \eqref{compact_en} is engineered 
in such a way that
this term linear in $F$
cancels in $de_n$,
leaving behind only one
term with two $F$'s.
This can be verfied with the help of 
some Schouten identities
and of the Bianchi identities
\be  \label{Bianchis}
0 = DF^{IJ} := dF^{IJ}
+ gA^{IK} \wedge F_{K}{}^J
+ g A^{JK} \wedge F^I{}_K \ , \quad
DDy^I = g F^{IJ} y_J \ .
\ee 
The relative coefficient
between the $j=1$
and $j=2$ terms in 
\eqref{compact_en}
is similarly engineered to
guarantee the cancellation
of the terms with two
$F$'s in $de_n$, leaving behind only a term with
three $F$'s.
For even $n$ this pattern of cancellations in $de_n$ continues until we reach
$de_n = 0$. For odd $n$,
the pattern of cancellations
proceeds until we get a $de_n$ that is purely horizontal, \emph{i.e.}~without any $Dy$
factors,
\begin{align}  \label{chi_eq}
\text{$n = 2m-1$:} \;\;\;\;
de_{2m-1} &= - \pi^* \chi_{2m}(E) \ , 
\\
\chi_{2m}(E) & = 
\frac{ g^{m} }{ 2^{m}  m! (2\pi)^m }  
\epsilon_{I_1 J_1 \dots I_m J_m} F^{I_1 J_1}
\wedge \dots \wedge F^{I_m J_m} 
=  {\rm Pf} \left( 
\frac{gF}{2\pi} \right)  \ . \nn
\end{align}

The quantity $e_n'$ that enters the
uplift formula \eqref{general_flux}
is constructed out of the canonical
global angular form $e_n$. More precisely,
depending on the specific case under examination
from Table \ref{table}, we go from
$e_n$ to $e_n'$ by truncating the
expansion of $e_n$ in powers of the field
strenghts $F^{IJ}$.  
In other words, $e_n'$ is obtained
by truncating the sum over $j$ in \eqref{compact_en}.
 The details of the truncation
are reported below for each case.



\subsection{Interpretation of the uplift formula for 
\texorpdfstring{$\hat {\mathcal F}_n$}{TEXT}
}

Let us comment on the form \eqref{general_flux}
of the uplift formula for the flux
$\hat \cF_n$ threading the sphere.

If the gauge fields of $SO(n+1)$ are set to zero,
and the scalar matrix $T_{IJ}$ is set to the identity,
the expression for $\hat \cF_n$ clearly reduces
to the volume form on the round $S^n$.
This is a closed form that represents 
the generator of the cohomology group
$H^n(S^n;\mathbb Z)$.
The physical flux is given by an integer times
this generator, by virtue of flux quantization.

Let us imagine to turn on the scalar fields $T_{IJ}$,
keeping the gauge fields zero for the moment.
Formula \eqref{general_flux} states
that, under the continuous deformation parametrized by $T_{IJ}$, 
$\hat \cF_n$ is shifted by an exact piece
of the form $d[(yTy)^{-1}\epsilon y (Ty)(dy)^{n-1}]$.
It follows that  the de Rham cohomology class
of $\hat \cF_n$ is unmodified.
This is to be expected, since a continuous
deformation cannot change the integral value of the flux
threading the sphere.
The detailed form of the exact deformation
of $\hat \cF_n$ when the scalars $T_{IJ}$ 
are turned on (but the gauge fields are zero)
was derived in \cite{Lee:2014mla} using
the notion of generalized 
parallelizability 
for spheres.
In section \ref{sec_integral} below
we offer a different argument
(similar to the considerations of \cite{Nastase:2000tu})
that determines both the functional form
and the numerical prefactor of the exact deformation.

We may alternatively start from the volume form
on the round $S^n$ and turn on the $SO(n+1)$ gauge fields,
keeping $T_{IJ}$ fixed to be the identity matrix.
In this case the total $D$-dimensional spacetime
should be regarded as a sphere fibration.
The `naked' volume form $\epsilon y (dy)^n$
is no longer a well-defined $n$-form in spacetime.
Rather, it must be promoted by means of the replacement
$dy^I \rightarrow Dy^I$.
The resulting $n$-form $\epsilon y (Dy)^n$
is indeed globally defined, yet it fails to be closed:
its derivative takes the form $\epsilon F (Dy)^{n-1}$.
The terms in the sum \eqref{general_e_prime}
with $j=1,2,\dots,j_{\rm max}$
can be interpreted as corrective terms:
as discussed in the previous section
around \eqref{Bianchis}, adding the term
with $j=1$ ensures that the non-closure
of $e'_n$ is of the form $\epsilon F^2 (Dy)^{n-3}$;
further adding the term with $j=2$ yields
a non-closure of the form 
$\epsilon F^3 (Dy)^{n-5}$; and so on.

Finally, let us comment on the case in which both the
gauge fields of $SO(n+1)$ and the scalars $T_{IJ}$
are turned on.
The uplift formula \eqref{general_flux},
based on all the explicit examples we have studied,
exhibits a particularly simple structure.
Indeed, the dependence on the scalar fields
$T_{IJ}$ is entirely confined inside a total derivative,
even after the gauge fields are activated.
Furthermore, the $(n-1)$-form
inside the total derivative is 
$(yTy)^{-1}\epsilon y (Ty)(Dy)^{n-1}$.
This quantity  is obtained from the exact deformation
when the gauge fields as zero
by means of the minimal replacement 
$dy^I \rightarrow Dy^I$ inside the total derivative.
A priori, additional terms could have
been added inside the total derivative after
turning on the gauge fields,
such as terms proportional to $\epsilon y (Ty) F (Dy)^{n-3}$.
In all examples we have studied, however,
we observe that such terms are not generated.
In section \ref{sec_bott}, in the case
of $D=11$ supergravity on $S^4$,
we offer a different perspective
on this fact, 
based on a formula of Bott and Cattaneo \cite{BottCattaneo}.

\subsection{Detailed uplift formulae}
\label{sec:uplift}

Let us now examine each case in Table \ref{table}
in turn.

\subsubsection{$D=11$ supergravity on $S^4$}
11d supergravity can be consistently
truncated on $S^4$ \cite{Pilch:1984xy,Nastase:1999cb,Nastase:1999kf}  to 7d maximal $SO(5)$ gauged supergravity
\cite{Pernici:1984xx}.
The flux that threads the sphere is clearly
the closed $G_4$ flux of 11d supergravity, 
\be
\hat \cF_4 \propto \hat G_4 \ .
\ee
Recall that, by definition,
$\hat \cF_4$ is rescaled in such a way as to integrate to 1 on $S^4$.
In this case there is no dilaton
in the higher-dimensional theory,
and thus no dilaton among the  modes
that are kept in seven dimensions.
The bosonic content of 7d maximal
$SO(5)$ gauged supergravity consists of the metric,
the $SO(5)$ gauge fields, 20 real scalars
parametrizing the coset $SL(5,\mathbb R)/SO(5)$,
and five 3-forms $c^I_3$, 
transforming in the vector representation
of $SO(5)$. The latter are the `extra' fields
reported in the last column of Table \ref{table}.

The complete uplift formulae for this
consistent truncation are given in \cite{Nastase:1999kf},
see also \cite{Cvetic:2000ah}.
By applying some Schouten identities, they can be 
recast in the following form,
\begin{align}
d\hat s^2_{11}  & = (yTy)^{1/3} \bigg[  ds^2_7 
+ g^{-2}\frac{1}{yTy} T^{-1}_{IJ} Dy^I Dy^J \bigg]  
 \ ,
\\
\hat \cF_4   & = e_4 + \frac{1}{\cV_4 4!} d \bigg[
\frac 43  \frac{1}{yTy}  \epsilon y (Ty) (Dy)^3
\bigg] + d(y_I c^I_3)   \ , \\
e_4 &= \frac{1}{\cV_4 4!} \bigg[ 
\epsilon y (Dy)^4
+ 2 g \epsilon y F (Dy)^2
+ g^2 \epsilon y F^2
\bigg]  \ .
\end{align}
We observe that in this case the quantity $e_4'$
that enters the uplift formula for $\hat \cF_4$
is exactly identified with the canonical
global angular form $e_4$. In other words,
the sum over $j$ in \eqref{compact_en} is not truncated.
Moreover, let us point out that the extra
three-forms $c_3^I$ do not enter the uplift
formula for the 11d metric,
and enter the uplift formula for the 11d four-form
flux in a very simple way, via a total derivative.

\subsubsection{$D=11$ supergravity on $S^7$}
11d supergravity can be consistently
truncated on $S^7$ \cite{deWit:1983vq,deWit:1984nz,deWit:1985iy,deWit:1986mz,deWit:1986oxb,Nicolai:2011cy,deWit:2013ija,Hohm:2013pua,Godazgar:2013dma,Hohm:2014qga,Godazgar:2015qia,Varela:2015ywx}
to 4d maximal $SO(8)$ gauged supergravity
\cite{deWit:1981sst,deWit:1982bul}.
The flux that threads the sphere is now
\be
\hat \cF_7 \propto   \hat * \hat G_4 \ .
\ee
The bosonic content of 4d maximal
$SO(8)$ gauged supergravity consists of the metric,
the $SO(8)$ gauge fields, 
and 70 real scalars
parametrizing the coset $E_{7(7)}/(SU(8)/\mathbb Z_2)$.
Out of these 70 scalars, 35 
have positive intrinsic parity
(proper scalars) and 35 have negative
intrinsic parity (pseudoscalars).
The 35 proper scalars parametrize the coset
$SL(8,\mathbb R)/SO(8)$. From this point of view,
the remaining 35 pseudoscalars
are regarded as `extra' fields in the
terminology of Table \ref{table}.

The complete uplift formulae for the
11d metric and four-form flux
for this consistent truncation
are given in \cite{Varela:2015ywx}, building on 
\cite{deWit:1983vq,deWit:1984nz,deWit:1985iy,deWit:1986mz,deWit:1986oxb,Nicolai:2011cy,deWit:2013ija,Hohm:2013pua,Godazgar:2013dma,Hohm:2014qga,Godazgar:2015qia}.
In order to verify the general formulae
\eqref{general_metric}, \eqref{general_flux},
the task at hand is: (i)
turn off the 35 pseudoscalars, so that
only the $SL(8, \mathbb R)/SO(8)$ scalars remain;
(ii) compute $\hat G_7$
by taking the Hodge dual
of the $\hat G_4$ flux given in 
\cite{Varela:2015ywx}.
Some steps of these computations are reported in
appendix \ref{app_S7}. The result reads
\begin{align}
d\hat s^2_{11}  & = (yTy)^{2/3} \bigg[  ds^2_4 
+ g^{-2}\frac{1}{yTy} T^{-1}_{IJ} Dy^I Dy^J \bigg]  
+ \text{(extra)} 
 \ ,
\\
\hat \cF_7   & = e_7' + \frac{1}{\cV_7 7!} d \bigg[
\frac 76 \frac{1}{yTy} \epsilon y (Ty) (Dy)^6
\bigg] + \text{(extra)}   \ , 
\label{G7_result}
\\
e_7' &= \frac{1}{\cV_7 7!}
\bigg[  
\epsilon y (Dy)^7
+ \frac 72 g \epsilon y F (Dy)^5
\bigg]  \ .
\end{align}
We notice that the 7-form $e_7'$
that enters the uplift formula
is now distinct from the canonical
global angular form $e_7$ given
by \eqref{compact_en}.
In the canonical $e_7$, we encounter terms
with up to three $F$'s, schematically
$e_7 \sim \epsilon y (Dy)^7 
+ \epsilon y F(Dy)^5 + \epsilon y F^2 (Dy)^3
+ \epsilon y F^3 Dy$. In this case
the base of the $S^7$ fibration
is 4d spacetime, and thus the $F^3$ term
vanishes for dimensional reasons.
The $F^2$ term in $e_7$, however,
survives on a 4d base space.
It constitutes the difference between
the canonical $e_7$ and its truncated counterpart
$e_7'$.

It is worth emphasizing that
in this consistent truncation
the extra fields (the 35 pseudo scalars)
enter both the uplift formula for the metric
and seven-flux
in non-trivial ways.
While the uplift formulae retaining
all 70 scalars
are known explicitly \cite{Varela:2015ywx},
we were not able to recast them
in a simple form in the same spirit as in \eqref{general_flux}.
We leave this problem for future research.

\subsubsection{$D=10$ type IIB supergravity on $S^5$}
10d type IIB supergravity can be consistently
truncated on $S^5$ \cite{Kim:1985ez,Cvetic:1999xp,Lu:1999bw,Cvetic:1999xx,Cvetic:2000eb,Khavaev:1998fb,Cvetic:2000nc,Pilch:2000ue,Cassani:2010uw,Liu:2010sa,Gauntlett:2010vu,Skenderis:2010vz,Baguet:2015sma}
to 5d maximal $SO(6)$ gauged supergravity
\cite{Gunaydin:1984qu,Pernici:1985ju,Gunaydin:1985cu}.
The flux that threads the sphere is the self-dual Ramond-Ramond
five-form flux $\hat F_5$.
To discuss the
uplift formula for $\hat F_5$,
we find it convenient
to introduce  a five-form $\hat \cF_5$
that satisfies
\be 
 \hat \cF_5
+ \hat * \hat \cF_5 \propto \hat F_5 \ .
\ee 
The bosonic content of 5d maximal
$SO(6)$ gauged supergravity consists of the metric,
the $SO(6)$ gauge fields, 42 real scalars
parametrizing the coset $E_{6(6)}/USp(8)$,
and a collection of twelve real two-form potentials
$B_2^{I \alpha}$, 
where
$I=1,\dots,6$ is a fundamental index of
$SL(6,\mathbb R)$ and $\alpha = 1,2$
is a fundamental index of $SL(2,\mathbb R)$.
Recall that $SL(6,\mathbb R) \times SL(2,\mathbb R)$ is a maximal
subgroup of $E_{6(6)}$.
The $SL(2,\mathbb R)$  factor is identified with the 
 global
$SL(2,\mathbb R)$ symmetry of
classical type IIB supergravity in ten dimensions.

The complete uplift formulae
for this consistent truncation
are given in \cite{Baguet:2015sma}.
In this work, however,
we restrict our attention
to a subset of the bosonic fields
of the 5d gauged supergravity,
namely those that are inert under
$SL(2,\mathbb R)$.
In ten dimensions,
the bosonic fields that are
singlets of $SL(2,\mathbb R)$
are the Einstein frame metric
and the five-form flux.
They form a 10d consistent bosonic
subsector of the full type IIB
supergravity.
This 10d bosonic theory
can be consistently truncated 
on $S^5$ to the bosonic theory
obtained from 5d maximal $SO(6)$
supergravity by discarding the fermions, the two-forms
$B_2^{I\alpha}$,
and carving out
a scalar submanifold  
$SL(6,\mathbb R)/SO(6)$
out of the full
$E_{6(6)}/USp(8)$ scalar
manifold.
The uplift formulae for this
truncation are given in \cite{Cvetic:2000nc}
and can be rearranged to take the form\footnote{ The quantity we denote $\hat \cF_5$
is denoted $\hat * \hat G_{(5)}$
in \cite{Cvetic:2000nc}, where it is given in
equation (4).}
\begin{align}
d\hat s^2_{10}  & = (yTy)^{1/2} \bigg[  ds^2_5 
+ g^{-2}\frac{1}{yTy} T^{-1}_{IJ} Dy^I Dy^J \bigg]  
 \ ,
\\
\hat \cF_5   & =  e_5' + \frac{1}{\cV_5 5!} d \bigg[
\frac 54 \epsilon y(Ty) (Dy)^4 
\bigg]    \ , 
\label{F5_result}
\\
e_5' &= \frac{1}{\cV_5 5!}
\bigg[  
\epsilon y (Dy)^5
+ \frac 52 g \epsilon y F (Dy)^3
\bigg]  \ .
\end{align}
We notice that the 5-form $e_5'$
that enters the uplift formula
is again distinct from the canonical
global angular form $e_5$ given
by \eqref{compact_en}.
In the canonical $e_5$, we encounter terms
with up to two $F$'s, schematically
$e_5 \sim \epsilon y (Dy)^5 
+ \epsilon y F(Dy)^3 + \epsilon y F^2 Dy$. 
The terms with two $F$'s is not identically
zero on external 5d spacetime,
yielding indeed  a non-zero difference between $e_5$ and
$e_5'$.

\subsubsection{Massive $D=10$ type IIA supergravity on $S^6$}
\label{sec_massive}
Massive type IIA supergravity can
be consistently truncated on $S^6$ to
maximal dyonically gauged $ISO(7)$ supergravity
\cite{Guarino:2015jca,Guarino:2015qaa, Guarino:2015vca}.
The bosonic fields of this 4d theory
consists of the metric,
electric gauge fields for the gauge group
$ISO(7) = SO(7) \ltimes \mathbb R^7 $ of dimension $21+7$,  70 real scalars
parametrizing the coset
$E_{7(7)}/(SU(8)/\mathbb Z_2)$,
as well as 7 magnetic gauge fields
and 7 2-form potentials.
The presence of magnetic gauge fields
and 2-form potentials is due to a
magnetic gauging of the $\mathbb R^7$
subgroup of $ISO(7)$ \cite{Guarino:2015qaa, deWit:2007kvg}.

The full uplift formulae
for this consistent truncation
are given in \cite{Guarino:2015vca}.
We want to study the terms in these formulae
that contain the 4d metric,
the electric gauge fields
associated to the $SO(7)$ subgroup
of $ISO(7)$, as well as a subset of the scalar
fields, consisting of 
the scalars $T_{IJ}$ in the coset $SL(7,\mathbb R)/SO(7)$, together with one extra 
real scalar $Y$.
All other bosonic fields
of the truncation as regarded as `extra'
in the terminology of Table \ref{table}.
In analogy to the consistent
truncations based on the bosonic
action \eqref{starting_action},
discussed below in section \ref{dilaton_cases},   we may think of $Y$ as  a 4d dilaton.
The precise embedding of
$T_{IJ}$ and $Y$
into $E_{7(7)}/(SU(8)/\mathbb Z_2)$
is described in appendix \ref{app_massive}.

Let us focus on the uplift formulae
for the 10d Einstein frame metric,
the 10d dilaton,
and the 10d flux that threads the $S^6$.
The latter is identified with
\be  \label{who_is_F6}
\hat \cF_6 \propto  e^{\frac 12 \hat \phi} \hat * \hat F_4 \ ,
\ee  
where $\hat F_4$ is the Ramond-Ramond 
4-form field strength.
Notice the appearance of a dilaton prefactor:
it is due to the fact that
the 10d equations of motion
take the form $d(e^{\frac 12 \hat \phi} \hat * \hat F_4) = \dots$.
The proportionality constant in 
\eqref{who_is_F6} is determined
by our choice of normalization
for $\hat \cF_6$ (which integrates to 1 on $S^6$).

The task at hand is to 
start from the uplift formulae
of \cite{Guarino:2015vca}, in the simplified
setting in which we set to zero the 4d
fields we are not keeping track of.
After obtaining
the expression for $\hat F_4$,
we can compute $e^{\frac 12 \hat \phi} \hat * \hat F_4$. 
Some details of this derivation
are reported in appendix \ref{app_massive}.
We find the following uplift formulae,
\begin{align}
d\hat s^2_{10}  & =Y^{\frac{3}{14}} (yTy)^{\frac 58}
\bigg[
ds^2_4
+ g^{-2} \frac{Y^{- \frac 27}}{yTy} T^{-1}_{IJ}
Dy^I Dy^J
\bigg]  
+ \text{(extra)}
\label{dyonic_metric}
 \ ,
\\
e^{4\hat \phi} & = 
Y^{\frac {20}{7}} (yTy)^{-1}
\times \text{(extra)}
\label{dyonic_dilaton}
\ , \\
\hat \cF_6   & =  e_6' + \frac{1}{\cV_6 6!} d \bigg[
\frac 65 \epsilon y(Ty) (Dy)^5 
\bigg] 
+ \text{(extra)}\ , 
\label{dyonic_flux}
\\
e_6' &= \frac{1}{\cV_6 6!}
\bigg[  
\epsilon y (Dy)^6
+ 3 g \epsilon y F (Dy)^4
\bigg]  \ . \label{dyonic_e6}
\end{align}
Interestingly, 
all $Y$ and $dY$ terms eventually
drop out from the expression for
$\hat \cF_6$, which only contains
the $SO(7)$ gauge fields
and the $SL(7,\mathbb R)/SO(7)$ scalars.
The quantity $e_6'$ is a truncation
of the canonical global angular form
$e_6$ as given in \eqref{compact_en}.
Indeed, on a 4d base space
the canonical $e_6$ contains three terms,
schematically
$e_6 \sim \epsilon y (Dy)^6 
+ \epsilon y F(Dy)^4 + \epsilon y F^2 Dy^2$,
while $e_6'$ lacks the $F^2$ term.

The powers of $yTy$ and $Y$ in 
\eqref{dyonic_metric}, \eqref{dyonic_dilaton}
are in agreement with the
general expressions \eqref{general_metric}, \eqref{general_dilaton}.
The sign $s$ is $+1$. This is due to the fact that
we are adopting the conventions of
\cite{Guarino:2015vca} for the action of massive $D=10$ type IIA
supergravity---see (A.1) therein. In particular, the kinetic term
for the Ramond-Ramond 4-form flux is of the form
$e^{+ \frac 12 \hat \phi} \hat F_4  \wedge \hat * \hat F_4$.
In terms of the dual flux $\hat \cF_6 \propto e^{\frac 12 \hat \phi } \hat * \hat F_4$, this term reads
schematically $e^{- \frac 12 \hat \phi} \hat \cF_6 \wedge 
\hat * \hat \cF_6$, from which we infer $s=+1$.

\subsubsection{Consistent truncations
based on \eqref{starting_action}}
\label{dilaton_cases}
These three cases
are discussed in 
 \cite{Cvetic:2000dm}, where the
 full uplift formulae are given.
The uplift formulae for the $D$-dimensional
metric and dilaton take the form\footnote{The
matrix $T$ in our expressions is always unimodular.
It corresponds to the matrix denoted
$\tilde T$ in  \cite{Cvetic:2000dm}.}
\begin{align}
d \hat s^2_D  = Y^{ c_1  } (yTy)^{ \frac{n-1}{D-2}  } 
\bigg[
ds^2_d + g^{-2} Y^{- \frac{2}{n+1}} \frac{1}{yTy}
T^{-1}_{IJ} Dy^I Dy^J
\bigg] \  ,
\quad e^{\frac{2s}{a}  \hat \phi}  =Y^{c_2} (yTy)^{-1}  \ .
\end{align}
The constants $c_1$, $c_2$ are as in \eqref{c1c2def}.
The sign $s$ is $+1$ for the reductions
on $S^2$, $S^3$ and $-1$ for the reduction 
on $S^{D-3}$. This can be seen from
\eqref{starting_action}.
For the reductions on $S^2$, $S^3$ the flux
that threads the sphere is the `electric' flux
$\hat F_p$, whose kinetic term
is of the form $e^{-a \hat \phi} \hat F_p \wedge \hat * \hat F_p$.
For the reduction on $S^{D-3}$ the flux threading the
sphere is the `magnetic' flux $e^{- a \hat \phi} \hat * \hat F_3$.
Written in terms of $e^{- a \hat \phi} \hat * \hat F_3$,
the kinetic term for $\hat F_p$ in \eqref{starting_action}
takes the form $e^{+a \hat \phi} ( e^{- a \hat \phi} \hat * \hat F_3   ) \wedge \hat * (e^{- a \hat \phi} \hat * \hat F_3)$,
from which we observe the anticipated
flip in sign in the exponent of the dilaton prefactor.
We adopt the same normalization for $Y$
as in \cite{Cvetic:2000dm}. The kinetic term
for $Y$ in the lower-dimensional action takes the form
\be  
S_{(d)} \supset \int - \frac {1}{2} c_5  Y^{-2} dY \wedge* dY  \ , \qquad
c_5 = \frac{4 (D-n-2)}{a^2 (D-2)(n+1)}   \ .
\ee 
The canonically normalized lower-dimensional
dilaton $\phi$ is related to $Y$ by
$Y = e^{\frac{1}{\sqrt {c_5}} \phi}$.

Let us now discuss the uplift formulae
for the flux threading the sphere, in each of the 
three cases. For the reduction on $S^2$, $\hat \cF_2 \propto \hat F_2$ with
\begin{align}
\hat \cF_2 &=  e_2 
+ \frac{1}{\cV_2 2!} d \bigg[ 
2 \frac{1}{yTy} \epsilon_{IJK} y^I (Ty)^J Dy^K
\bigg]
\ , \\  
e_2 &= \frac{1}{\cV_2 2!} \epsilon_{IJK} y^I ( 
Dy^J \wedge Dy^K 
+ g F^{JK}
) \ .
\end{align} 
The notation $\cV_n$
was introduced in \eqref{unit_sphere}.
This result is given in equation
(55) of \cite{Cvetic:2000dm}
in terms of the quantity
denoted $T^{ij}$ there,
which corresponds to
$Y^{1/3} T_{IJ}$ in our notation. Remarkably,
all occurrences of $Y$
and its derivatives drop
away from the expression
for $\hat \cF_2$.
We also 
notice that $e_2$ is the 
canonical global angular form \eqref{compact_en}, satisfying $de_2 =0$.

For the reduction on $S^3$
we find similarly $\hat \cF_3 \propto \hat F_3$ with
\begin{align}
\hat \cF_3 & =  e_3
+ \frac{1}{\cV_3 3!} d\bigg[
\frac 32 \frac{1}{ yTy} \epsilon_{IJKL} y^I (Ty)^J Dy^K \wedge Dy^L
\bigg] + f_3 \ , \\
e_3 & = \frac{1}{\cV_3 3!} 
\epsilon_{IJKL} y^I 
\left(
Dy^J \wedge Dy^J \wedge Dy^K
+ \frac 32 g F^{JK} \wedge Dy^L
\right) \ .
\end{align}
The quantity $f_3$ is the field strength of a 2-form
potential in the $d=D-3$
dimensional theory.
This 2-form potential is an extra field needed for the
consistency of the truncation,
see Table \ref{table}.
Up to normalization, $f_3$
is the same as the field
$F_{(3)}$ in the notation
of \cite{Cvetic:2000dm}.
We observe once more that
 all factors of $Y$ and $dY$ drop from the 
uplift formula for the flux
threading the sphere.
 The quantity
$e_3$ coincides in this case
with the canonical \eqref{compact_en}. Indeed,
the derivative of $e_3$ is purely horizontal,
\be  \label{de3horizontal}
de_3 = - \frac{g^2}{8(2\pi)^2} \epsilon_{IJKL} F^{IJ} \wedge F^{KL} \ .
\ee 
Recall that the
$D$-dimensional flux
$\hat \cF_3$ is closed.
Its closure holds
because the extra 2-form
potential with field strength $f_3$
has a non-trivial Bianchi
identity involving the
$SO(4)$ gauge fields \cite{Cvetic:2000dm},
with $df_3$
exhibiting the same
structure as the RHS of \eqref{de3horizontal}.
In fact, 
in computing $d\hat \cF_3$
we verify  an exact cancellation between
$de_3$ and $df_3$.

We finally turn to the reduction
on $S^{D-3}$. 
The flux threading the sphere is in this case given by
\be 
\hat \cF_{D-3} \propto e^{-a \hat \phi} \hat * \hat F_3 \ .
\ee 
We have defined the dual
flux $\hat \cF_{D-3}$
by including a suitable power
of the $D$-dimensional dilaton.
This
is motivated by the fact that
the EOM for $\hat F_3$ derived from the
action \eqref{starting_action} reads
\be 
d(e^{- a \hat \phi}  \hat * \hat F_3)= 0 \ .
\ee 
Thus, in $D$ dimensions
$\hat F_3$ is closed
off-shell and 
$\hat \cF_{D-3}$
is closed on-shell.

The uplift formula for $\hat \cF_{D-3}$
is given in (47) of \cite{Cvetic:2000dm},
which can be written as
\begin{align} \label{magnetic_case}
\hat \cF_{D-3} &   = 
e'_{D-3}
+ \frac{1}{\cV_{D-3} (D-3)!}
d \bigg[ 
\frac{D-3}{D-4}
\frac{1}{yTy}
\epsilon y (Ty) (Dy)^{D-4}
\bigg] \ , \\
e'_{D-3}
& = \frac{1}{\cV_{D-3} (D-3)!} \bigg[  \epsilon y (Dy)^{D-3}
+ \frac{D-3}{2}
\epsilon y F (Dy)^{D-5}
\bigg] \ .
\end{align}
We are adopting a  
compact notation analogous to that in 
\eqref{shorthand}.
In analogy with  the other cases discussed
in this section, we verify that all
occurrences of $Y$ and $dY$
drop out from the uplift
formula for $\hat \cF_{D-3}$. In this case,
we find generically a truncated version
$e'_{D-3}$ of the canonical
global angular form
$e_{D-3}$, because 
the sum over $j$ in
\eqref{compact_en} is truncated
after the first term.
(For $D \le 6$
the sum over $j$
in the canonical global angular form $e_{D-3}$
stops anyway at 1;
thus for $D \le 6$
we actually have 
$e'_{D-3} = e_{D-3}$.)
The fact that
we truncate the sum over $j$
at $j=1$ implies that, for general $D$,
$de'_{D-3}$ 
consists of one term quadratic in $F$,
of the schematic form
$\epsilon F^2 (Dy)^{D-6}$.
The lower-dimensional theory,
however, lives in $d=D-(D-3)=3$ dimensions,
implying that $de'_{D-3} = 0$ for dimensional
reasons.
We this conclude that the form of the uplift
formula \eqref{magnetic_case} is such that
$d \hat \cF_{D-3}=0$ holds identically,
without use of the 3d equations of motion.
The closure of $\hat F_3$, 
on the other hand, holds after using
them.


\section{Reduction of the action and exact flux deformation}
\label{sec_integral}

In this section we give an argument that explains
the form of the exact piece in the uplift formula
\eqref{general_flux} for the flux that threads the $n$-sphere.
The argument is based on the derivation of the 
scalar potential of the lower-dimensional model 
from the $D$-dimensional action integrated on $S^n$ (see also \cite{Nastase:2000tu}). 

\subsection{Models without dilaton}

Let us first consider the cases in which,
in the  
$D$-dimensional theory, the kinetic term for the flux
threading the sphere has no dilaton prefactor.
The relevant terms in the $D$-dimensional
action are the Einstein-Hilbert term
and the kinetic term for $\hat \cF_n$,
\be  \label{EH_and_kin}
S_{(D)} =   \int \bigg[  
\hat R \hat * 1 
- \frac{1}{2 g^2_{\cF}} \hat \cF_n \wedge \hat * \hat \cF_n
\bigg]  \ .
\ee 
Recall that, in our conventions,
$\hat \cF_n$ is normalized to integrate to $1$ on $S^n$.
As a result, its kinetic term is not necessarily
canonically normalized in $D$-dimensions,
but rather comes with a coupling constant
$g^{-2}_\cF$. The following argument does not 
depend on the value of $g^{-2}_\cF$.
We also notice that we have suppressed the
overall $D$-dimensional Newton's constant
from the action \eqref{EH_and_kin}.

Our goal is to integrate \eqref{EH_and_kin} on
$S^n$ to derive  couplings in the $d$-dimensional
effective action. More precisely, we seek to
determine the potential for the scalars $T_{IJ}$.
To this end, we may ignore the terms with the gauge fields
of $SO(n+1)$, and further consider the simpler case
in which the profile for the $T_{IJ}$ scalars
in $d$ dimensions is constant,
\be   \label{simpler_case}
A_{IJ} = 0 \ , \qquad dT_{IJ} = 0 \ . 
\ee

To perform the dimensional reduction
of the Einstein-Hilbert term in \eqref{EH_and_kin} we need 
an ansatz for the $D$-dimensional metric,
capturing the contributions of the $T_{IJ}$ scalars.
We use the ansatz
\be   \label{ansatz_with_T}
d\hat s^2_D = (yTy)^{b_1} \Big[  ds^2_d
+ g^{-2} (yTy)^{b_2}  T^{-1}_{IJ} dy^I dy^J 
\Big]  \ ,
\ee  
where $b_1$, $b_2$ are constant parameters.
In due course, they will be fixed
to their expected values
$b_1= (n-1)/(D-2)$, $b_2 = -1$, cfr.~\eqref{general_metric}.
The $D$-dimensional Ricci scalar
and volume element of the metric 
\eqref{ansatz_with_T} can now be evaluated 
by means of a straightforward computation.
We keep track of terms
that contribute to the $d$-dimensional
scalar potential and to the $d$-dimensional
Einstein-Hilbert term.
Further details of the derivation are reported in 
appendix \ref{app_no_dilaton}.
The result reads
\begin{align} \label{EH_reduction}
\int_{S^n} \hat R \hat * 1
& \supset 
g^{-n}  \sqrt{-g_d} R[g_d] \int_{S^n} d^n \xi  \sqrt{\accentset{\circ}{g}}
(yTy)^{\frac{1}{2} b_1 (D-2)+\frac{1}{2}b_2
   n+\frac{1}{2}}
\nn \\
& + g^{2-n}  \sqrt{-g_d}
\int_{S^n} d^n \xi  \sqrt{\accentset{\circ}{g}}
(yTy)^{\frac{1}{2} b_1 (D-2)+\frac{1}{2} b_2
   (n-2)+\frac{1}{2}} \mathcal G \ .
\end{align}
Here $\xi^m$ are local coordinates on $S^n$,
 $g_d$, $R[g_d]$ are the determinant 
and Ricci scalar of the $d$-dimensional metric,
and 
$\accentset{\circ}{g}$ is the determinant of the
metric   on the round unit sphere.
We have introduced the shorthand notation
\begin{align}    \label{cG_expression}
\cG & =  \frac{( {\rm Tr} \, T )^2}{ yTy} - \frac{{\rm Tr} \, T^2}{yTy}
- \frac{2 \, ({\rm Tr} \, T) \, (yT^2y)}{(yTy)^2} + \frac{2 \,  yT^3y }{(yTy)^2} 
  + \cK \bigg[
 \frac{(yT^2 y)^2}{(yTy)^3}
 - \frac{(yT^3 y)}{(yTy)^2}
  \bigg] \ ,
\end{align}
where the constant $\cK$ is given
in terms of $D$, $n$, $b_1$, $b_2$ by
\be
\cK = -2 b_2 b_1 (D-2) (n-1)
- b_1^2 (D-2)
   (D-1)-b_2^2 (n-2) (n-1) \ .
\ee  
In the quantities $yT^2y$ and $y T^3y$
we are suppressing $SO(n+1)$ indices
as in \eqref{index_contractions}.

To perform the integration of 
the $\hat \cF_n$ kinetic term
in \eqref{EH_and_kin} over the $n$-sphere, we need an 
ansatz capturing the dependence of $\hat \cF_n$ on the scalars
$T_{IJ}$. Since we are working under the simplifying assumptions
\eqref{simpler_case}, $\hat \cF_n$ must be proportional to the
volume form of the round metric on $S^n$, where the proportionality
factor may depend on $y^I$ and $T_{IJ}$,
\be \label{f_definition}
\hat \cF_n = \frac{1}{n!} f(y,T) \epsilon_{I_0 I_1 \dots I_n} y^{I_0} 
dy^{I_1} \wedge \dots \wedge dy^{I_n} \ .
\ee  
The reduction     is reported in appendix
\ref{app_no_dilaton}. The result reads 
\begin{align} \label{flux_reduction_res_temp}
\int_{S^n} \hat \cF_n \wedge \hat * \hat \cF_n
& = g^n \sqrt{- g_d} \int_{S^n}  d^n \xi 
\sqrt{\accentset{\circ}{g}}  (yTy)^{\frac{b_1 D}{2}-b_1 n - \frac{b_2n}{ 2} 
 - \frac 12 }  f(y,T)^2\ .
\end{align}

In the $d$-dimensional action the Einstein-Hilbert term
should not be multiplied by a non-trivial function 
of $T_{IJ}$. This leads us to set the exponent of $yTy$
in the first line of \eqref{EH_reduction} to zero,
\be   \label{first_lin_rel}
\frac{1}{2} b_1 (D-2)+\frac{1}{2}b_2
   n+\frac{1}{2} = 0 \ .
\ee
For the integrand in the second line of \eqref{EH_reduction}
we then find the quantity $(yTy)^{-b_2} \cG$.
The matrix $T_{IJ}$ is unimodular,
but
it is convenient to regard the quantities
$yTy$ and $\cG$
as functions of an arbitrary non-singular symmetric
matrix. In this sense, we can formally
consider their behavior under a constant
rescaling $T_{IJ} \rightarrow \lambda T_{IJ}$.
We see that $yTy$ and $\cG$ are homogeneous of degree 1 in $T$,
implying that the combination
$(yTy)^{-b_2} \cG$ has degree $1-b_2$.
We know, however, that the scalar potential of the
$d$-dimensional theory
should be a quadratic function of $T$.
This is known to be the case  for all examples
in Table \ref{table} with a `no' in the dilaton column.
These considerations lead us to set $1- b_2 = 2$.
In combination with  \eqref{first_lin_rel} this fixes $b_1$ and $b_2$  
to their expected values,
\be   \label{fix_b1_b2}
b_1 = \frac{n-1}{D-2} \ , \qquad b_2 = -1 \ .
\ee 
The overall integrand
in the second line of \eqref{EH_reduction}
then becomes $(yTy) \cG$.

To proceed, we observe that 
we are free to add to 
$(yTy) \cG$ any total divergence on $S^n$,
without affecting the value of the integral.
Exploiting this freedom,   \eqref{EH_reduction}
can be brought to the form (see appendix \ref{app_no_dilaton}
for details)
\begin{align}  
& \int_{S^n} \hat R \hat * 1
  \supset 
g^{-n}  \cV_n \sqrt{-g_d} R[g_d] 
 + g^{2-n}  \sqrt{-g_d}
\int_{S^n} d^n \xi  \sqrt{\accentset{\circ}{g}} \, \cI \ , \nn \\
& \cI   = 
2 \bigg[ \frac{ yT^2y }{ yTy } 
- \frac {4 - k}{8}   {\rm Tr} \, T 
 \bigg]^2
 - \bigg( 1 + \frac k 2  \bigg) \, {\rm Tr} \, (T^2)
+ \bigg( \frac 12 + \frac k 4 - \frac{k^2}{32}  \bigg) \, ({\rm Tr} \, T)^2
 \ ,  \label{EH_term}
\end{align}
where we have introduced the shorthand notation
\be \label{k_eq}
k := \frac{(n-1)(D-n-1)}{D-2}  - 2  \ .
\ee  
Note that inside the integrand $\cI$
all dependence on $y^I$ is now collected
into a perfect square.
Moreover,
  using \eqref{fix_b1_b2}, \eqref{k_eq}
we can rewrite \eqref{flux_reduction_res_temp} in the form
\begin{align} \label{flux_reduction_res}
\int_{S^n} \hat \cF_n \wedge \hat * \hat \cF_n
& = g^n \sqrt{- g_d} \int_{S^n} d^n \xi  
\sqrt{\accentset{\circ}{g}}  (yTy)^{2+k}  f(y,T)^2\ .
\end{align}

Upon integrating $\cI$ against the volume form
$\sqrt{\accentset{\circ}{g}}$,
we obtain a function of $T_{IJ}$
which represents the contribution of the $D$-dimensional
Einstein-Hilbert term to the $d$-dimensional scalar potential.
The resulting function of $T_{IJ}$
is manifestly homogeneous of degree 2 under
$T_{IJ} \rightarrow \lambda T_{IJ}$. It is not, however, a quadratic function. 
In fact,
it is not a rational function of the entries of $T_{IJ}$.
This is due to the contribution
from the square $[ \frac{ yT^2y }{ yTy } 
- \frac {4 - k}{8}   {\rm Tr} \, T ]^2$
in the integrand.
This is point is clarified
in appendix \ref{app_integral_example}.
Non-quadratic contributions to the scalar potential 
must drop
out when 
the Einstein-Hilbert contribution is added to 
\eqref{flux_reduction_res}.
Comparison of \eqref{EH_term}, \eqref{flux_reduction_res}
immediately suggests a simple and natural mechanism
to achieve this goal:
the term $[ \frac{ yT^2y }{ yTy } 
- \frac {4 - k}{8}   {\rm Tr} \, T ]^2$
in  $\cI$ is cancelled
against $(yTy)^{2+k} f(y,T)^2$
in \eqref{flux_reduction_res} at the level of the \emph{integrand}.
The remaining terms in $\cI$
are a $y$-independent  quadratic function of~$T$.

For the rest of this section,
the aforementioned cancellation
between $[ \frac{ yT^2y }{ yTy } 
- \frac {4 - k}{8}   {\rm Tr} \, T ]^2$
and $(yTy)^{2+k} f(y,T)^2$
is our working assumption.
Therefore, we set
\be \label{f_identification}
f(y,T) = 2 g_\cF g^{1-n}  (yTy)^{-\frac{k}{2}}\bigg[ \frac{ yT^2y  }{(yTy)^2} - \frac{4-k}{8} \frac{{\rm Tr} \, T }{yTy}\bigg] \ .
\ee


A necessary condition for the
identification \eqref{f_identification} to hold
is that the integral
of the RHS on $S^n$ with measure
$\sqrt{\accentset{\circ}{g}}$ 
be independent of $T$.
Indeed, the integral of the LHS
must be independent of $T$,
because turning on $T$
must 
deform $\hat \cF_n$ by an exact piece
(the
quantized flux through $S^n$
cannot be modified by a continuous
deformation).
In appendix \ref{app_integrals_and_T}
we prove that a necessary condition
for the integral of the RHS of 
\eqref{f_identification}
to be independent of $T$ is
\be  
k = 0 \ .
\ee  
Crucially, this condition is also sufficient
to ensure that the integral of $f(T,y)$ is independent
of $T$. This follows from the observation that
\be   \label{why_it_is_sufficient}
 \epsilon y (dy)^n + d\bigg[
\frac{n}{n-1} \frac{1}{yTy} \epsilon y (Ty) (dy)^{n-1}
\bigg] = - \frac{2}{n-1}\bigg[ \frac{ yT^2y  }{(yTy)^2} - \frac 12 \frac{ {\rm Tr} \, T  }{yTy}
\bigg]  \epsilon y (dy)^n  \ .
\ee  
The integral of the LHS over $S^n$
is manifestly independent of $T$ by virtue
of Stokes' theorem.

To summarize, 
our assumptions are:
(1) turning on $T$ deforms the flux 
by an exact piece;
(2) the perfect square in the Einstein-Hilbert and the contribution from the kinetic term for $\hat \cF_n$
cancel against each other at the level
of the $S^n$ integrand.
Under these assumptions,
we determine uniquely
the form of $f(y,T)$
and therefore
of $\hat \cF_n$ via \eqref{f_definition},
up
to the overall normalization.
The latter is readily
fixed by requiring 
$\int_{S^n} \hat \cF_n = 1$.
The result is
\be  
\hat \cF_n  = \frac{1}{\cV_n n!}  \epsilon y (dy)^n +
\frac{1}{ \cV_n n!}  d\bigg[
\frac{n}{n-1} \frac{1}{yTy} \epsilon y (Ty) (dy)^{n-1}
\bigg]  \ ,
\ee  
in perfect agreement with
\eqref{general_flux}
and \cite{Lee:2014mla}.
Both the functional form
of the exact deformation of the flux,
as well as its  numerical coefficient
relative to the undeformed 
flux, are determined\footnote{More precisely, the $(n-1)$-form 
inside the total derivative
is completely fixed up
to shifts by closed $(n-1)$-forms,
which clearly would not
modify the expression for $\hat \cF_n$.
}.

Remarkably, 
in the process of our derivation
we have found the condition
$k=0$, which
is a constraint on the possible values of $n$, $D$. In fact,
this condition 
selects precisely the values for $n$, $D$ for which a consistent truncation is possible with the field content we are considering. Setting \eqref{k_eq} equal to zero, one finds that the only integer solutions with $n<D$ are given by
\be
(n,D) =(1,2) , (4,11)  ,  (5,10) , (7,11)    \ .
\ee
While the first of these would describe reduction on a circle from 2d to 1d, the remaining three correspond to familiar consistent truncations. This is a striking result. Based on the requirement that the $d$-dimensional scalar potential be quadratic in the scalars $T_{IJ}$ and that the flux be deformed by an exact piece, we are able to back out the possible dimensions of the sphere and the lower-dimensional model. 
Similar conclusions were drawn in \cite{Nastase:2000tu}
based on slightly different arguments.\footnote{Compared
to \cite{Nastase:2000tu}, our argument is based
on a cancellation
at the level of the \emph{integrand}
on $S^n$. Thus, we do not have to evaluate
explicitly $S^n$ integrals of rational
functions of $y$, $T$, which   in general
yield non-rational functions of the entries of $T$,
see the example \eqref{integral_example} in appendix
\ref{app_no_dilaton}.}

\subsection{Models with dilaton}

The analysis of the previous section can be repeated
for the cases in which the kinetic term for
$\hat \cF_n$ comes with a non-trivial dilaton prefactor.
The relevant terms in the $D$-dimensional action
are now 
\be   \label{EH_and_kin_and_dil}
S_{(D)} =   \int \bigg[  
\hat R \hat * 1 
- \frac{1}{2 g^2_{\cF}} e^{-sa \hat \phi} \hat \cF_n \wedge \hat * \hat \cF_n
- \frac 12 d \hat \phi \wedge \hat *  d \hat \phi 
\bigg]  \ .
\ee  
where $a$ is a positive constant 
and $s$ is a sign.
Below we will see how the value 
\eqref{a_def} for $a$ is selected.
Notice that the $D$-dimensional dilaton
$\hat \phi$ has a canonically normalized kinetic term.
Our goal is to determine the scalar
potential of the $d$-dimensional theory.
As a result, we may work under the simplifying assumptions
\be  \label{simpler_with_dilaton}
A^{IJ} = 0 \ , \qquad dT_{IJ} =0   \ , \qquad
dY = 0 \ ,
\ee  
where $Y$ encodes the $d$-dimensional dilaton.

The metric ansatz must be modified,
to take into account   $Y$.
We adopt the following modification of \eqref{ansatz_with_T},
\be  \label{ansatz_with_T_and_Y}
d\hat s^2_D = (yTy)^{b_1} Y^{b'_1} \Big[  ds^2_d
+ g^{-2} (yTy)^{b_2} Y^{b'_2} T^{-1}_{IJ} dy^I dy^J 
\Big]  \ ,
\ee  
where $b_1$, $b_2$, $b'_1$, $b'_2$
are constant parameters, to be fixed momentarily.
The reduction of the $D$-dimensional Einstein-Hilbert
term can be performed analogously to the previous section.
The result reads
\begin{align} \label{EH_reduction_with_dil}
\int_{S^n} \hat R \hat * 1
& \supset 
g^{-n} Y^{\frac{b_1'D}{2} + \frac{b_2'n}{2} - b_1' }  \sqrt{-g_d} R[g_d] \int_{S^n} d^n \xi   \sqrt{\accentset{\circ}{g}}
(yTy)^{\frac{1}{2} b_1 (D-2)+\frac{1}{2}b_2
   n+\frac{1}{2}}
\nn \\
& + g^{2-n} Y^{\frac{b_1'D}{2} + \frac{b_2'n}{2} - b_1' -b_2' } \sqrt{-g_d}
\int_{S^n} d^n \xi  
\sqrt{\accentset{\circ}{g}} (yTy)^{\frac{1}{2} b_1 (D-2)+\frac{1}{2} b_2
   (n-2)+\frac{1}{2}} \mathcal G \ ,
\end{align}
with the same $\cG$ as in \eqref{cG_expression}.

The reduction of the kinetic term for $\hat \cF_n$
is also analogous to the previous case.
We adopt the same ansatz for the flux as \eqref{f_definition},
except that now we allow $f$ to be a function of $Y$, too,
\be \label{f_definition_with_Y}
\hat \cF_n = \frac{1}{n!} f(y,T,Y) \epsilon_{I_0 I_1 \dots I_n} y^{I_0} 
dy^{I_1} \wedge \dots \wedge dy^{I_n} \ .
\ee  
We also use the following ansatz for the $D$-dimensional dilaton,
\be  \label{dilaton_ansatz}
e^{s \hat \phi} = (yTy)^{b_3} Y^{b_3'} \ ,
\ee 
where $b_3$, $b_3'$ are real constant parameters.
Reducing the kinetic term for $\hat \cF_n$ yields
\begin{align} \label{flux_reduction_dilaton_res_temp}
& \int_{S^n} e^{-sa \hat \phi}\hat \cF_n \wedge \hat * \hat \cF_n
 =  \\
 & \quad = g^n
Y^{\frac{b_1'D}{2} - b_1'n - \frac{b_2'n}{2} - a b_3'}
\sqrt{- g_d} \int_{S^n}  d^n \xi 
\sqrt{\accentset{\circ}{g}}  (yTy)^{\frac{b_1 D}{2}-b_1 n - \frac{b_2n}{ 2} 
 - \frac 12 - a b_3 }  f(y,T,Y)^2\ . \nn 
\end{align}

Finally, we have to take into account the
contributions to the $d$-dimensional
scalar potential
originating from the
kinetic term for $\hat \phi$.
We have
\begin{align}   \label{dilaton_kin_temp}
- \frac 12 \int_{S^n} d \hat \phi\wedge \hat * 
d \hat \phi  & = 
2 g^{2-n} b_3^2 
Y^{\frac{b_1'D}{2} - b_1' + \frac{b_2'n}{2} - b_2' } 
 \times   \\
& \times  \sqrt{-g_{d}} \int_{S^n} d^n \xi 
\sqrt{ \accentset{\circ}{g}} \,
(yTy)^{\frac{b_1D}{2}-b_1 + \frac{b_2n}{2}-b_2+\frac 12 } \, 
 \, \bigg[
 \frac{(yT^2 y)^2}{(yTy)^3}
 - \frac{(yT^3 y)}{(yTy)^2}
  \bigg]   \ .  \nn
\end{align}

In order to get a canonical Einstein-Hilbert
term in $d$ dimensions from \eqref{EH_reduction_with_dil},
we set
\be   \label{wanna_be_canonical}
\frac{1}{2} b_1 (D-2)+\frac{1}{2}b_2
   n+\frac{1}{2} = 0 \ , \qquad 
\frac{b_1'D}{2} + \frac{b_2'n}{2} - b_1' = 0  \ .
\ee 
Exactly as in the previous 
section, we can examine how the second
line of \eqref{EH_reduction_with_dil}
scales with $T$, impose an overall
scaling of degree $2$, and thus obtain
another linear relation
between $b_1$ and $b_2$.
The net result is to fix these parameters
as before, see \eqref{fix_b1_b2}.
Based for instance on the results of \cite{Cvetic:2000dm},
we know that the scalar potential,
written as a function of unimodular $T$
and $Y$, is also homogeneous in rescalings of $Y$,
with degree $2/(n+1)$.
We may alternatively regard this condition
as a way of (partially) fixing 
ambiguities related to redefinitions of $Y$.
If we demand that the second line of 
\eqref{EH_reduction_with_dil}
be homogeneous of degree $2/(n+1)$ in $Y$
 we get a second linear relation
in $b_1'$, $b_2'$, which combined with 
\eqref{wanna_be_canonical} gives us
\be   \label{fix_other_bs}
b'_1 = \frac{2n}{(D-2)(n+1)}  \ , \qquad
b'_2 = - \frac{2}{n+1} \ .
\ee  
These values guarantee that 
\eqref{ansatz_with_T_and_Y}
matches with our previous expression
\eqref{general_metric}.

Having fixed $b_1$, $b_2$ according
to \eqref{fix_b1_b2}, we can mimick the same steps
as in the previous section.
By adding a suitable total divergence on $S^n$,
we can eliminate the $yT^3y$ terms 
from the sum of \eqref{EH_reduction_with_dil}
and \eqref{dilaton_kin_temp},
obtaining the simpler form
\begin{align} \label{EH_plus_dilaton}
& \int_{S^n} \bigg[  
\hat R \hat * 1 
- \frac 12 d \hat \phi \wedge \hat * d \hat \phi
\bigg]
= g^{-n}  \cV_n \sqrt{-g_d} R[g_d] 
 + g^{2-n} Y^{\frac{2}{n+1}} \sqrt{-g_d}
\int_{S^n}
d^n \xi  \sqrt{\accentset{\circ}{g}} \, \tilde  \cI \ , \nn \\
& \tilde \cI   = 
2 \bigg[ \frac{ yT^2y }{ yTy } 
- \frac {4 - \tilde k}{8}   {\rm Tr} \, T 
 \bigg]^2
 - \bigg( 1 + \frac{ \tilde k}{ 2}  \bigg) \, {\rm Tr} \, (T^2)
+ \bigg( \frac 12 + \frac {\tilde k}{ 4} - \frac{\tilde k^2}{32}  \bigg) \, ({\rm Tr} \, T)^2
 \ ,  
\end{align}
where we have introduced
\be   \label{tildek_def}
\tilde  k = \frac{(n-1)(D-n-1)}{D-2}  - 2
+ 2 b_3^2 \ .
\ee  
Remarkably,
the functional form
of the integrand in \eqref{EH_plus_dilaton}
is exactly the same as in \eqref{EH_term},
up to the replacement 
$k \rightarrow \tilde k$.

Making use of \eqref{fix_b1_b2}, \eqref{fix_other_bs},
and \eqref{tildek_def},
the contribution \eqref{flux_reduction_dilaton_res_temp}
from the flux kinetic term
takes the form
\begin{align} \label{flux_reduction_dilaton_res}
& - \frac{1}{2g_\cF^2} \int_{S^n} e^{-sa \hat \phi}\hat \cF_n \wedge \hat * \hat \cF_n
  = \nn \\
  & \qquad = - \frac{1}{2g_\cF^2} g^n
Y^{\frac{2 n (D-n-1)}{(D-2) (n+1)}-a b_3'}
\sqrt{- g_d} \int_{S^n}  d^n \xi 
\sqrt{\accentset{\circ}{g}}  (yTy)^{
2 + \tilde k - a b_3 - 2 b_3^2
}  f(y,T,Y)^2\ .
\end{align}

Once again, a natural simple mechanism
suggests itself to 
eliminate non-rational functions of $T$
originating from the square in $\tilde \cI$
in \eqref{EH_plus_dilaton}: a cancellation 
against $f^2$ at the level of the integrand.
We thus proceed assuming
\be \label{f_identification_dil}
f(y,T,Y) = 2 g_\cF g^{1-n}
Y^{
\frac{a b_3'}{2}+\frac{D-D
   n+n^2+n-2}{(D-2) (n+1)}
}
(yTy)^{-\frac{\tilde k}{2} + \frac 12 a b_3
+ b_3^2}\bigg[ \frac{ yT^2y  }{(yTy)^2} - \frac{4-\tilde k}{8} \frac{{\rm Tr} \, T }{yTy}\bigg] \ .
\ee
As in the previous section,
we assume that $f(y,T,Y)$ originates
from an exact deformation of the 
round volume form on $S^n$.
A necessary condition for the identification
\eqref{f_identification_dil}
to
be possible is therefore that the $S^n$ integral
of the RHS with measure $\sqrt{ \accentset{\circ}{g} }$
be independent of $T$ and $Y$.
Independence on $Y$ is easily
established by tuning the value of $b_3'$,
\be  
b_3' = \frac{2 (n-1) (D-n-2)}{a (D-2) (n+1)} \ .
\ee  
This value guarantees that
\eqref{dilaton_ansatz} agrees with our previous
expressions
\eqref{general_dilaton}, \eqref{c1c2def}.
In appendix \ref{app_integrals_and_T}
we study  
when the integral of \eqref{f_identification_dil}
becomes independent of $T$,
as we vary the parameters $\tilde k$, $\tfrac 12 a b_3  + b_3^2$.
We identify two possibilities.

The first possibility is 
 to set
\be   \label{dilaton_conditions}
\tilde k = 0 \ , \qquad
\frac 12 a b_3  + b_3^2  = 0 \ . 
\ee  
For these values of 
$\tilde k$, $\tfrac 12 a b_3  + b_3^2$
the RHS of 
\eqref{f_identification_dil}
becomes the same as in the case without dilaton
\eqref{f_identification} with $k=0$.
We already know from the discussion around 
\eqref{why_it_is_sufficient} that the integral
 is independent of $T$,
for any symmetric
matrix $T$, unimodular or not.



As observed above, in the cases without a dilaton
the condition $k=0$ fixes the allowed values of $D$,
$n$. In the present context, 
we can solve 
the conditions \eqref{dilaton_conditions} 
in terms of the parameters $a$ and $b_3$.
We find
\be  
a = - 2b_3 \ , \qquad a^2 = 4 
-\frac{2 (n-1) (D-n-1)}{D-2} \ .
\ee 
This is in agreement with \eqref{a_def}:
we have thus found a different argument
that singles out the value of $a^2$  for which the
truncation is possible
\cite{Cvetic:2000dm}.
Moreover,
the relation $b_3 = -a/2$
ensure the compatibility of \eqref{dilaton_ansatz}
 with our previous formula
\eqref{general_dilaton}.

Let us now turn to the second
possible choice for $\tilde k$, $\tfrac 12 a b_3  + b_3^2$ 
that  renders the integral
of \eqref{f_identification_dil} independent of $T$,
\be   \label{second_tildek}
\tilde k = 
\frac{4 (n+1)}{n+3}  \ , \qquad
\frac 12 a b_3  + b_3^2 = -\frac{(n-1) (n+1)}{2 (n+3)} \ .
\ee  
This case is qualitatively different from
\eqref{dilaton_conditions}: as discussed in appendix
\ref{app_integrals_and_T},
if we choose \eqref{second_tildek}
the integral of \eqref{f_identification_dil}
becomes a constant times $(\det T)^{- \frac 12}$,
hence a constant for unimodular $T$,
but not for a generic symmetric $T$.
The values \eqref{second_tildek}
are not compatible with the uplift formulae
recorded earlier and checked against the literature.
The interpretation of \eqref{second_tildek}
seems more elusive and is left for future
investigation.


\section{Bott-Cattaneo formula and $S^4$ truncation}
\label{sec_bott}

In the previous section 
we have studied a simplified setting
in which the $SO(n+1)$ gauge fields 
are set to zero, and the $T$ scalars
to a constant.
We have demonstrated that 
$\hat \cF_n$ must contain
an exact term given by the derivative of the $(n-1)$-form
$(yTy)^{-1} \epsilon y (Ty) (dy)^{n-1}$, with the appropriate numerical
prefactor.
Let us now suppose we turn on the
$SO(n+1)$ gauge fields,
and we relax the assumption that $T$
is constant.
Gauge invariance requires
the replacement
$dy \rightarrow Dy$,
leading to establish
that the exact deformation
of $\hat \cF_n$
must contain the term
$(yTy)^{-1} \epsilon y (Ty) (Dy)^{n-1}$.
A priori, 
additional terms
might be generated inside the exact
deformation, proportional to
$F$ and/or $DT$.
The explicit analysis
of the uplift formulae in
all examples collected in table
\ref{table}, however,
shows that this does not happen.
While we cannot furnish a first-principle proof of this fact,
we can point out an interesting
connection with some results
of Bott and Cattaneo \cite{BottCattaneo}
regarding the fiber integrals of
global angular forms for
even-dimensional spheres.

\paragraph{Reminder on the Bott-Cattaneo formula.}
Recall from section \ref{GAF_review}
that the canonical global angular form
$e_n$ 
for an $n$-sphere with $n$ even
is a closed form
in the total space of an $S^n$ fibration
over a base space $B$.
In this setting, 
we can consider the $k$th power of 
$e_n$ and fiber-integrate it along the
$S^n$ directions to obtain
a closed $n(k-1)$-form on the base space $B$. The result of this operation
is the content of the Bott-Cattaneo formula \cite{BottCattaneo},
\be   \label{BC_formula}
\int_{S^{2m}} (e_{2m})^{2s+2} = 0 \ , \qquad
\int_{S^{2m}} (e_{2m})^{2s+1}
= 2^{-2s} (p_m)^s \  , \qquad s = 0,1,2,\dots 
\ee   
Here $n=2m$ is the dimension of the sphere, and $p_m$ denotes the
$m$th Pontryagin \emph{form} constructed 
with the
curvature $F^{IJ}$ of the $SO(n+1)$ gauge fields.
This is a polynomial in $F^{IJ}$
of degree $m$,
  a closed $4m$-form on the base
space $B$ with integral periods.
Its  cohomology class
corresponds to an element of
$H^{2m}(B;\mathbb Z)$.

\paragraph{Application to the reduction
of $D=11$ supergravity on $S^4$.}
In this case, the flux $\hat \cF_4$ that threads the $S^4$
is the  $\hat G_4$ flux
of 11d supergravity.
The 11d action
contains a Chern-Simons term
$\hat C_3 \hat G_4 \hat G_4$.
We are interested in analyzing
the 7d couplings that are generated
by fiber-integrating this Chern-Simons
term along $S^4$.
Following \cite{Freed:1998tg,Harvey:1998bx},
this is most easily performed
by regarding 11d spacetime as the boundary
of an auxiliary 12d space,
which is an $S^4$ fibration over
an 8d base $B_8$.
The physical 7d spacetime
is the boundary of $B_8$.
We can consider 
the 
formal 12-form
$\hat G_4^3$, do the fiber-integration to get an 8-form on $B_8$,
and consider 
its restriction to the 7d boundary of
$B_8$.

We know that $\hat G_4 \propto \hat \cF_4$ takes the form
\be  
\hat \cF_4 = e_4 + d\omega_3 \ ,
\ee
where $\omega_3$ is a globally-defined
3-form. We thus can write
\be   \label{expand_F4cube} 
\int_{S^4} \hat \cF_4^3
= 
\int_{S^4} \Big( e_4^3 + d\omega_{11}   \Big) 
= \int_{S^4} e_4^3
+ d \int_{S^4} \omega_{11}
= \tfrac 14 p_2  + d \int_{S^4} \omega_{11} \ .
\ee 
In the previous expressions we have defined the 11-form
\be  \label{omega11_def}
\omega_{11} = 3 e_4^2 \wedge \omega_3
+ 3 e_4 \wedge \omega_3 \wedge d\omega_3
+ \omega_3 \wedge (d\omega_3)^2  \ ,
\ee 
and we have used the fact that
fiber-integration and exterior
derivative commute.
In the final step 
of \eqref{expand_F4cube} we have applied 
the Bott-Cattaneo formula \eqref{BC_formula}.

The RHS of \eqref{expand_F4cube}
is an 8-form that encodes 
topological couplings in the 7d action. 
The   term with $p_2$ 
in the 8d bulk 
corresponds to a  Chern-Simons
form on the 7d boundary.
Such coupling is indeed
found
in the action of 7d maximal
$SO(5)$ gauged supergravity
\cite{Pernici:1984xx}.

The term $d \int_{S^4} \omega_{11}$
in \eqref{expand_F4cube}
is an exact deviation
from the Bott-Cattaneo formula.
At the level of cohomology classes,
the Bott-Cattaneo relation
persists, but it is no longer
valid at the level of differential
forms.
The term $d \int_{S^4} \omega_{11}$
in the 8d bulk
can generate   additional
topological
couplings  $\int_{S^4} \omega_{11}$
on the 7d boundary.

We may now observe the following. We know that the consistent truncation selects the following form
for $\omega_3$,
\be   \label{good_omega3}
\omega_3 = \frac{1}{\cV_4 4!} \frac 43  \frac{1}{yTy} \epsilon y (Ty) (Dy)^3  \ .
\ee  
Plugging this $\omega_3$ into
\eqref{omega11_def}, we verify
that the 7-form $\int_{S^4} \omega_{11}$ vanishes identically.\footnote{In fact,
this is true as soon as $\omega_3$
is proportional to $Dy^I \wedge Dy^J \wedge Dy^K$,
since it is easy to verify that, in this case, no term in $\omega_{11}$
has four factors $Dy$, which is necessary
to yield a non-zero result
upon fiber-integrating over $S^4$.}
Thus, 
the Bott-Cattaneo
relation persist at the level
of differential forms, and no additional topological couplings are generated in the 7d action.

This remarkable property is generically
lost if we consider a more general 
$\omega_3$. For example,
we can imagine adding a term
with one field strength $F$, of the form
\be   \label{new_omega3}
\omega_3 = \frac{1}{\cV_4 4!} \bigg[ \frac 43  \frac{1}{yTy} \epsilon y (Ty) (Dy)^3
+ \xi_2 \frac{1}{yTy} \epsilon y (Ty) F Dy
\bigg] \ ,
\ee  
where $\xi_2$ is a constant coefficient.
Notice that the new term goes to zero
if we set $T = \mathbb I$.
We can repeat the computation of
$\int_{S^4} \omega_{11}$
with the new $\omega_3$ in \eqref{new_omega3}.
The result is non-zero,
of the schematic form
\be   \label{cS_coupling}
\int_{S^4} \omega_{11} \sim \cS(T)_{I_1 \dots I_8}
DT^{I_1 I_2} \wedge F^{I_3 I_4}
\wedge F^{I_5 I_6}
\wedge F^{I_7 I_8} \ .
\ee  
Here $\cS(T)_{I_1 \dots I_8}$
stands for a tensor
with 8 free $SO(5)$ indices,
constructed with $T$.
A non-zero $\int_{S^4} \omega_{11}$
implies that the Bott-Cattaneo
formula is no longer valid at the level
of forms.

To summarize,
the consistent truncation
selects an exact deformation
with precisely three $Dy$ legs.
Such a deformation is non-generic,
in that it ensures that
the Bott-Cattaneo relation
persists at the level of differential
forms. 
A deviation from a term
with three $Dy$'s
generically induces
topological terms in the 7d
action of the form 
\eqref{cS_coupling}.\footnote{Such terms have four derivatives.
This is more than the three
derivatives in the 
supergravity action built 
by the Noether procedure 
from the Einstein-Hilbert term.
(The term with three
derivatives is the Chern-Simons
term $d^{-1}p_2$.)
In a maximal supergravity theory,
the first higher-derivative
corrections are expected to
emerge at eight derivatives.
These naive considerations
suggest that terms such as
\eqref{cS_coupling} might be incompatible
with supersymmetry.
}


\section{Outlook}
\label{sec_outlook}

In this work we have 
made progress towards a
unified approach to 
consistent truncations on spheres,
centered around 
the 
classical geometric notion
of 
global angular form.
The latter is a mathematical
object defined 
for an $n$-sphere bundle over a base space.
In applications to consistent
truncations,
the base space is the spacetime of the lower-dimensional theory,
and the sphere fiber is the
internal space used
in the compactification.

Our main results are equations
\eqref{general_metric}-\eqref{general_flux}.
They
describe universal features
of the consistent Kaluza-Klein ans\"atze 
that are common to all the 
cases listed in Table \ref{table}.
In particular,
our  formulae
capture the contributions of
the $SO(n+1)$ gauge fields,
the $SL(n+1,\mathbb R)/SO(n+1)$ scalars describing the naive deformation space of $S^n$,
and of the dilaton (when present).
The ansatz \eqref{general_flux}
for the flux 
threading the sphere 
takes a  
particularly compact form,
with the $SL(n+1,\mathbb R)/SO(n+1)$ scalars
entering only via an exact 
shift. We have also shown
how this exact shift is determined
by computing the scalar potential
of the lower-dimensional model
by dimensional reduction,
and imposing mild constraints on its
functional dependence on the scalars.

An important problem that deserves
further investigation is
the analysis of the `extra fields'
that are required for consistency
of the truncation in some
cases (see last column of Table \ref{table}).
In particular,
it is desirable to identify
a  way
of detecting the necessity of extra fields based on the 
uplift formulae \eqref{general_metric}-\eqref{general_flux},
without  relying on the structure of
supersymmetry multiplets
in the lower-dimensional
supergravity theory,
and without
performing a full explicit check
that the lower-dimensional equations
of motion imply the higher-dimensional ones.
Progress in this direction  would be particularly useful
in light of  applications
to consistent truncations with
less-than-maximal supersymmetry. 
In this context, 
matter supermultiplets 
can appear in the lower-dimensional model,
making it more challenging
to identify which modes
should be retained for consistency
of the reduction.
It would also be beneficial
to try to uncover general patterns
in the way in which 
extra fields, when present,
modify the uplift formulae
\eqref{general_metric}-\eqref{general_flux},
with the aim to learn lessons to
be applied to more general setups
with less supersymmetry.

In our approach,
it is convenient to regard
the construction of the consistent
Kaluza-Klein ansatz for the flux as a two-step procedure.
The first step is the identification
of the flux with the relevant (truncated)
global angular form $e'_n$,  see~\eqref{eq:start}. At this stage
we capture the $SO(n+1)$ gauge fields related to the isometry of the sphere. As observed in the introduction, this stage
is sufficient for the determination
of the Chern-Simons terms
in the lower-dimensional
supergravity (for the $AdS_7$ and $AdS_5$ cases).
The second step is the analysis
of the effect of turning
on scalar deformations.

The analog of the starting point
\eqref{eq:start} is available
for more general setups,
in which the internal space
is not necessarily a sphere
and the $d$-dimensional
theory has less than maximal supersymmetry. 
In fact, 
we may consider
replacing $S^n$ with 
a compact internal space $M_n$,
which is then fibered
over a base space $B_d$,
$M_n \hookrightarrow X_{d+n} \rightarrow B_d$.
If $M_n$ has a continuous
isometry group $G$,
we turn on $G$-connections
on the bundle $X_{d+n}$. They correspond
to gauge fields with  gauge group
$G$ in the $d$-dimensional
theory formulated on $B_d$.
In this context, an important task
is the identification of 
globally defined forms
on the total space $X_{d+n}$
which reduce to a given closed form on $M_n$
if restricted to the fiber directions.\footnote{Mathematically,
this is  related to  
promoting  an element of $H^\bullet(M_n;\mathbb R)$ to a $G$-equivariant
cohomology class in $H^\bullet_G(M_6;\mathbb R)$.
This process can be obstructed \cite{WU1993381}.
The physical interpretation of such obstructions
involves the St\"uckelberg mechanism for
$p$-form gauge fields \cite{PhysRevD.89.086007}
(see also \cite{Bah:2021mzw,Bah:2021hei} 
for an analysis of this phenomenon
in 4d QFTs engineered with M5-branes).
}
In \cite{Bah:2019rgq,Bah:2020jas,Bah:2020uev} this problem has been studied
systematically
to compute
anomalies
in brane engineering and holography (see also \cite{Hosseini:2020vgl}).
The methods and results of 
\cite{Bah:2019rgq,Bah:2020jas,Bah:2020uev} could also provide 
starting points for the construction
of consistent Kaluza-Klein ans\"atze.\footnote{For instance,
$M_6$ could be an $S^4$ bundle over
a Riemann surface. See \cite{MatthewCheung:2019ehr,Cassani:2020cod}
for consistent truncations
with such an internal space.}









\acknowledgments

We thank C.~Hull and D.~Waldram for useful discussions.
RM is supported in
part by ERC Grant 787320-QBH Structure and by ERC Grant 772408-Stringlandscape. 
The work of VVC was supported by the Alexander von Humboldt Foundation via a Feodor Lynen fellowship.
The work of FB is supported
by the Simons Collaboration Grant on Categorical Symmetries.  PW is supported in part by NSF grant PHY-2112699.

\appendix
\addtocontents{toc}{\protect\setcounter{tocdepth}{1}}

\section{Analysis of uplift formulae}

\subsection{Conventions for round spheres}
\label{app_spheres}

The unit round $S^n$ is defined by the equation
\be 
\delta_{IJ} y^I y^J = 1
\ee 
in $\mathbb R^{n+1}$ with Cartesian coordinates $y^I$, $I=1,\dots,n+1$.
Throughout this appendix
we raise/lower $I$ indices with $\delta$.
We use $\xi^m$, $m=1,\dots,n$
for local coordinates on $S^n$,
with $\partial_m =\partial/\partial\xi^m$.
The round metric
on $S^n$
reads
\be  \label{round_metric_def}
\accentset{\circ}{g}_{mn}
= \delta_{IJ} \partial_m y^I \partial_n y^J 
\ .
\ee 
It admits the Killing vectors
\be  \label{round_Killing}
\mathcal K_{IJ}^m = \accentset{\circ}{g}^{mn} ( y_I 
\partial_{n} y_J 
- y_J 
\partial_{n} y_I ) \ ,
\ee 
where $\accentset{\circ}{g}^{mn}$
is the inverse of $\accentset{\circ}{g}_{mn}$.
A useful identity is
\be  \label{useful_identity}
\accentset{\circ}{g}^{mn} \partial_m y^I \partial_n y^J = \delta^{IJ} - y^I y^J \ .
\ee 
The quantities $y^I$, regarded
as a set of $n+1$ scalar functions on
$S^n$, satisfy 
\be  \label{conformal_scalars}
\accentset{\circ}{\nabla}_m \accentset{\circ}{\nabla}_n y^I = - y^I 
\accentset{\circ}{g}_{mn} \ ,
\ee 
where $\accentset{\circ}{\nabla}_m$
denotes the Levi-Civita 
connection associated to
$\accentset{\circ}{g}_{mn}$.
As a consequence of \eqref{conformal_scalars},
the quantities $y^I$
are eigenfunctions of the Laplacian
constructed with the round metric,
\be
\accentset{\circ}{g}^{mn} \accentset{\circ}{\nabla}_m \accentset{\circ}{\nabla}_n y^I
= -n  y^I \ .
\ee

\subsection{Formulae for Hodge stars}
\label{app_Hodge}

Let us consider a $D$-dimensional metric
of the form
\be  \label{some_metric}
d\hat s^2_D = e^{2A} ds^2_d + g_{mn} D\xi^m D\xi^n \ ,
\ee 
where $\xi^m$, are local 
 coordinates on
the $n$-sphere
with constrained coordinates $y^I$.
The 1-forms $D \xi^m$
are related to the 1-forms
$Dy^I$ in \eqref{Dy_def} by
\be  \label{Dxi_def}
Dy^I = \partial_{m} y^I D\xi^m \ , 
\qquad
D\xi^m = d\xi^m - \tfrac 12 g A^{IJ} \cK_{IJ}^m \ ,
\ee 
with $\mathcal K^m_{IJ}$ as in 
\eqref{round_Killing}.
The relation \eqref{useful_identity}
is useful in verifying the
compatibility between
 \eqref{Dxi_def}
and \eqref{Dy_def}.
We allow
the quantities 
$e^{2A}$ and for $g_{mn}$ 
in \eqref{some_metric} to depend both on the external coordinates and on the coordinates on $S^n$.

Suppose $\alpha$ is a 
$q$-form  with legs along the external spacetime
$d s^2_d$ only. The following identity holds,
\begin{align} \label{Hodge_identity}
& \hat * ( \alpha \wedge Dy^{I_1} \wedge \dots
\wedge Dy^{I_p})
 = (-)^{p(d-q)} e^{(d-2q)A} \frac{\sqrt g}{\sqrt {\accentset{\circ}{g}}} (* \alpha) \wedge 
 \\
& \qquad \qquad \wedge \frac{1}{(n-p)!} \epsilon_{J_0 J_1 \dots J_p K_1 \dots K_{n-p}} y^{J_0} Q^{I_1 J_1} \dots Q^{I_p J_p}
Dy^{K_1} \wedge \dots \wedge Dy^{K_{n-p}} \ . \nn
\end{align}
The symbol $\hat *$ on the LHS denotes the Hodge
star with respect to the metric 
\eqref{some_metric}, while $*$ on the RHS is the Hodge star
with respect to the external metric $ds^2_d$.
The quantities $g$, $\accentset{\circ}{g}$
are the determinants of 
$g_{mn}$, $\accentset{\circ}{g}_{mn}$, respectively.
The symmetric tensor $Q^{IJ}$ is defined as
\be  \label{Q_def}
Q^{IJ} = g^{mn} \partial_m y^I \partial_n y^J \ ,
\ee 
where $g^{mn}$ is the inverse of $g_{mn}$
in \eqref{some_metric}.

The   formula 
\eqref{Hodge_identity}
takes a simpler form
if specialized to 
a metric $g_{mn}$
of the form 
\be  \label{main_application}
g_{mn} =  e^{2A'}  T^{-1}_{IJ} \partial_m y^I \partial_n y^J \ ,
\ee 
where $A'$ is an arbitrary function
of external and internal coordinates
and  $T_{IJ}$ is a symmetric unimodular matrix, independent on the coordinates on $S^n$.
In this case,
the inverse $g^{mn}$ of $g_{mn}$
can be written in closed form in terms of the Killing vectors \eqref{round_Killing},
\be  
g^{mn} = e^{-2A'}  \frac{1}{2(yTy)} T^{IJ} T^{KL} \cK^m_{IK}
\cK^m_{JL} \ ,
\ee  
as it may be verified using \eqref{useful_identity}.
Plugging this expression into the $Q$ tensor
\eqref{Q_def} and using again \eqref{useful_identity}
we obtain
\be 
Q^{IJ} = e^{-2A'}   \bigg[
T^{IJ} - \frac{(Ty)^I (Ty)^J}{yTy}
\bigg] \ .
\ee  
The determinant of $g_{mn}$ in \eqref{main_application}
can also be written in closed form
(see \emph{e.g.}~\cite{Lee:2014mla}),
\be 
\det g_{mn} = e^{2A'n} (yTy) \det \accentset{\circ}{g}_{mn} \  , \qquad
\frac{\sqrt g}{\sqrt {\accentset{\circ}{g}}}
= e^{A'n}(yTy)^{1/2} \ ,
\ee 
where we recalled that $T$ is unimodular.

Let us close this section by sketching the derivation of \eqref{Hodge_identity}.
We can write the metric \eqref{some_metric} in the form
\be 
d\hat s^2_D  = e^{2A} ds^2_d + \delta_{\bar m \bar n}
e^{\bar m}  e^{\bar n} \ , \qquad
e^{\bar m} = e^{\bar m}{}_p D\xi^p \ ,
\ee 
where a bar denotes a flat index of $S^n$ and the
vielbein components $e^{\bar m}{}_p$
satisfy
$\delta_{\bar m \bar n} e^{\bar m}{}_p
e^{\bar n}{}_q = g_{pq}$.
Note that $e^{\bar m}{}_p$ depends in general on both internal and external coordinates.
By using the inverse $e_{\bar m}{}^p$
of $e^{\bar m}{}_p$ we can write
\be 
\alpha \wedge Dy^{I_1} \wedge \dots Dy^{I_p}
= \partial_{m_1} y^{I_1} \dots 
\partial_{m_p} y^{I_p} 
e_{\bar n_1}{}^{m_1} \dots
e_{\bar n_p}{}^{m_p} \alpha \wedge e^{\bar n_1} \wedge 
\dots \wedge e^{\bar n_p} \ .
\ee 
The Hodge star of 
$ \alpha \wedge e^{\bar n_1} \wedge 
\dots \wedge e^{\bar n_p}$
splits into the wedge product of
$*\alpha$
and $*_g(e^{\bar n_1} \wedge 
\dots \wedge e^{\bar n_p})$,
where $*_g$ is the Hodge star with respect to $g_{mn}$. In the splitting, 
we generate a
suitable power of the warp factor $A$, and the sign factor $(-)^{p(d-q)}$ from a reordering of indices.

We may now evaluate the quantity
\begin{align}
& \partial_{m_1} y^{I_1} \dots 
\partial_{m_p} y^{I_p} 
e_{\bar n_1}{}^{m_1} \dots
e_{\bar n_p}{}^{m_p} *_g(e^{\bar n_1} \wedge 
\dots \wedge e^{\bar n_p})
= \nn \\
& = \frac{1}{(n-p)!} \partial_{m_1} y^{I_1} \dots 
\partial_{m_p} y^{I_p} 
e_{\bar n_1}{}^{m_1} \dots
e_{\bar n_p}{}^{m_p} \epsilon^{\bar n_1 \dots \bar n
\bar r_1 \dots \bar r_{n-p}}
e_{\bar r_1} \wedge \dots \wedge e_{\bar r_{n-p}} \ .
\end{align}
Here the epsilon symbol with flat indices
takes values in  the set $\{0,\pm 1\}$.
Making use of $e^{\bar m} = e^{\bar m}{}_p D\xi^p$
we can also recast the same quantity
in a way that does not make explicit
reference to the vielbein,
\begin{align}
& \partial_{m_1} y^{I_1} \dots 
\partial_{m_p} y^{I_p} 
e_{\bar n_1}{}^{m_1} \dots
e_{\bar n_p}{}^{m_p} *_g(e^{\bar n_1} \wedge 
\dots \wedge e^{\bar n_p})
= \nn \\
& = \frac{1}{(n-p)!} \partial_{m_1} y^{I_1} \dots 
\partial_{m_p} y^{I_p} 
g^{m_1 s_1} \dots g^{m_p s_p}
\epsilon_{s_1 \dots s_p r_1 \dots r_{n-p}} D\xi^{r_1}
\wedge \dots \wedge D\xi^{r_{n-p}} 
 \ . \label{intermediate}
\end{align}
Now the epsilon tensor with lower curved indices
stands for the volume form of the metric $g_{mn}$ and therefore 
takes values in the set $\{ 0, \pm \sqrt g \}$.

The final step is to relate the epsilon tensor
of the metric 
$g_{mn}$ to that of the round metric $\accentset{\circ}{g}_{mn}$, and write the latter
in terms of the contrained coordinates $y^I$,
\be 
\epsilon_{m_1 \dots m_n} =
\frac{\sqrt g}{\sqrt{ \accentset{\circ}{g}}} \accentset{\circ}{\epsilon}_{m_1 \dots m_n}
= \frac{\sqrt g}{\sqrt {\accentset{\circ}{g}}}
\epsilon_{I_0 I_1 \dots I_n } y^{I_0} \partial_{m_1} y^{I_1} \dots \partial_{m_n} y^{I_n} \ . 
\ee 
Plugging this into \eqref{intermediate}
and making use of $\partial_m y^I D\xi^m = Dy^I$
and the definition \eqref{Q_def} of the tensor $Q^{IJ}$, we arrive at the desired formula
\eqref{Hodge_identity}.

\subsection{$D=11$ supergravity on $S^7$}
\label{app_S7}

The complete uplift formulae
for this consistent truncation
are given in \cite{Varela:2015ywx},
including all scalars
from the coset $E_{7(7)}/(SU(8)/\mathbb Z_2)$.
The latter 
enter the uplift formulae
via the symmetric matrices
$\mathcal M_{MN}$, $\mathcal I_{\Lambda \Sigma}$,
$\mathcal R_{\Lambda \Sigma}$.
Here 
$M,N = 1,\dots, 56$
are indices of the $\mathbf{56}$ 
of $E_{7(7)}$,
$\Lambda, \Sigma = 1,\dots,28$
are indices of the $\mathbf {28}$
of $SL(8,\mathbb R)$. A $\Lambda$ index is equivalent to pair of antisymmetrized
fundamental indices $[IJ]$ of $SL(8,\mathbb R)$.
In terms of $SL(8,\mathbb R) \subset E_{7(7)}$,
a lower $M$ index splits into a lower $\Lambda$ index, and an upper
$\Lambda$ index.
The matrix $\mathcal M_{MN}$
can be written in block form as
\be 
\mathcal M_{MN} =
\begin{pmatrix}
\mathcal M_{\Lambda \Sigma}
& \mathcal M_\Lambda{}^\Sigma  \\
\mathcal M^{\Lambda }{}_\Sigma
& \mathcal M^{\Lambda \Sigma}
\end{pmatrix} \ .
\ee 
The four blocks of $\mathcal M_{MN}$
and the matrices 
$\mathcal I_{\Lambda \Sigma}$,
$\mathcal R_{\Lambda \Sigma}$
are related by (see \emph{e.g.}~\cite{Guarino:2015qaa})
\be 
\mathcal M_{\Lambda \Sigma} = - (\mathcal I 
+ \mathcal R \mathcal I^{-1} \mathcal R)_{\Lambda \Sigma} \ ,  \quad 
\mathcal M_{\Lambda}{}^{\Sigma} = 
\mathcal M^{\Sigma}{}_{\Lambda}=(\mathcal R \mathcal I^{-1})_{\Lambda}{}^{\Sigma}  \  , \quad
\mathcal M^{\Lambda \Sigma} = 
- (\mathcal I^{-1})^{\Lambda \Sigma}
 \ .
\ee 
On the RHS of the previous relations,
$\Lambda$ indices are contracted according to matrix multiplication.\footnote{When we identify a $\Lambda$ index with an antisymmetric pair $[IJ]$, we insert a factor $1/2$ for each contracted pair.
For instance,
$(\mathcal R \mathcal I^{-1})_{[I_1 I_2]}{}^{[J_1 J_2]} = \tfrac 12 \mathcal R_{[I_1 I_2][K_1 K_2]} (\mathcal I^{-1})^{[K_1 K_2][J_1 J_2]}$
and $(\mathcal I \mathcal I^{-1})_{[I_1 I_2]}{}^{[J_1 J_2]} = \tfrac 12 \mathcal I_{[I_1 I_2] [K_1 K_2]} (\mathcal I^{-1})^{[K_1 K_2][J_1 J_2]} = \mathbb I_{[I_1 I_2]}{}^{[J_1 J_2]} :=  2\delta_{[I_1}^{J_1} \delta_{I_2]}^{J_2}$.
}
The matrix $\mathcal I_{\Lambda \Sigma}$ is invertible
and negative definite.
If we set to zero the 35 pseudoscalars, and only retain
the 35 proper scalars, we have
\be  \label{AdS4_no_pseudo}
\mathcal R_{\Lambda \Sigma} = 0 \ , \qquad
\mathcal I_{\Lambda \Sigma} = \mathcal I_{[IJ][KL]}
= -   T^{-1}_{IK}
 T^{-1}_{JL} 
+    T^{-1}_{IL}
T^{-1}_{JK} \ .
\ee 
The quantity $T^{-1}_{IJ}$ is
the inverse of the symmetric unimodular matrix
parametrizing the 35 proper scalars
of $SL(8,\mathbb R)/SO(8)$.

The specialization
\eqref{AdS4_no_pseudo}
implies considerable 
simplifications in
the general uplift formulae
of \cite{Varela:2015ywx}.
The expressions for the internal metric
and the warp factor
are given in (24) and (26) of 
\cite{Varela:2015ywx}
and can be unpacked with the help
of the expressions
(C.3) in \cite{Guarino:2015qaa}
for the generators
of $SL(8,\mathbb R) \subset E_{7(7)}$.
The result is
\be
d \hat s^2_{11} =
(yTy)^{2/3} ds^2_4
+
g^{-2} (yTy)^{-1/3}  T^{-1}_{IJ} D y^I D y^J 
\ ,
\ee 
with $(yTy)$ and $Dy^I$ as in \eqref{general_metric},
\eqref{Dy_def}.

The uplift formula for
$\hat G_4$ is (27) in \cite{Varela:2015ywx}
(where the 4-form flux is denoted
$\hat F_{(4)}$).
Thanks to \eqref{AdS4_no_pseudo},
the term $dA$ drops away.
The quantities $\mathcal H_{(4)}^{IJ}$,
$\mathcal H_{(3) I}{}^J$,
$\tilde \cH_{(2)IJ}$ in 
(27) of \cite{Varela:2015ywx} can be evaluated
using (16) in  \cite{Varela:2015ywx}.
The components of the $X$ tensor
that appears there
can be found for instance in
(C.9), (C.10) of \cite{Guarino:2015qaa}
(for the $SO(8)$ gauging, plug
$\theta_{AB} = \delta_{AB}$ and $\xi^{AB} = 0$ in those equations).
We arrive at the following
expression for the 4-form flux
(in our notation),
\begin{align}
\hat G_4  &= - \frac{1}{2g^2} (*F)^{JL} \wedge Dy^I \wedge Dy^K  T^{-1}_{IJ}  T^{-1}_{KL}
- \frac 1g T^{JK} y_K Dy^I \wedge  *DT^{-1}_{IJ}
\nn \\
& + g {\rm vol}_4 \Big[
({\rm Tr} T) (yTy) - 2 (yTTy)
\Big] \ .
\end{align}
Here  $yTTy = y_I T^{IJ} T_{JK} y^K$
and ${\rm vol}_4$ is the volume form
of the external metric $ds^2_4$.
We may now compute the Hodge star of $\hat G_4$
with the help of the identities collected
in section \ref{app_Hodge}
and verify
that the resulting $\hat G_7$ takes the form 
\eqref{G7_result} given in the main text.

\subsection{Massive $D=10$ type IIA supergravity on $S^6$}
\label{app_massive}

In this subsection,
we use $A$, $B = 1,\dots,8$
for fundamental indices of
$SL(8,\mathbb R)$,
reserving
$I$, $J=1,\dots,7$ to fundamental
indices of $SL(7,\mathbb R)$ or $SO(7)$.
The full set of scalar fields
in four dimensions is again parametrized
in terms of the matrices $\cM_{MN}$,
$\cI_{\Lambda \Sigma}$, $\cR_{\Lambda \Sigma}$
as above.
Now, we identify $\Lambda$ indices
with antisymmetrized pairs $[AB]$.
In a first step, we freeze to zero all pseudoscalar modes, so that
\be  \label{massive_no_pseudo}
\mathcal R_{\Lambda \Sigma} = 0 \ , \qquad
\mathcal I_{\Lambda \Sigma} = \mathcal I_{[AB][CD]}
= -   \cT^{-1}_{AC}
 \cT^{-1}_{BD} 
+    \cT^{-1}_{AD}
\cT^{-1}_{BC} \ .
\ee 
The quantity $\cT$ is a symmetric unimodular
$8 \times 8$ matrix.
Notice that \eqref{massive_no_pseudo} is nothing but 
\eqref{AdS4_no_pseudo}, written in a slightly
different notation
that is better suited to
discuss the problem at hand.

Next, let us perform the index splitting
$A \rightarrow (I,8)$.
We 
 restrict the matrix $\cT^{AB}$  to take the following block-diagonal form,
\be 
\cT^{AB}  =
\begin{pmatrix}
\cT^{IJ} & \cT^{I8} \\
\cT^{8J} & \cT^{88}
\end{pmatrix}
= \begin{pmatrix}
Y^{1/7} T^{IJ} & 0 \\
0 & Y^{-1}
\end{pmatrix} \ .
\ee 
Here $Y$ is a positive real scalar
and $T^{IJ}$ is the symmetric
unimodular matrix that parametrizes
$SL(7,\mathbb R)/SO(7)$.

The uplift formulae for
the metric and the dilaton
can be extracted 
from (3.14), (3.18), (3.23) in \cite{Guarino:2015vca}.
Since we only keep a subset of the scalars,
the quantities denoted $A_m$, $B_{mn}$,
$A_{mnp}$ there are all zero.
One may then readily reproduce 
the formulae 
\eqref{dyonic_metric},
\eqref{dyonic_flux} given in the main text.

Next, we turn to the evaluation of the Ramond-Ramond 4-form field strength.
It is denoted $\hat F_{(4)}$ in 
\cite{Guarino:2015vca}
and it is given by (A.4) therein
in terms of the $p$-form potentials
of type IIA.
Using (3.12) in \cite{Guarino:2015vca},
keeping in mind that $A_m = B_{mn} = A_{mnp} = 0$ for us,
we confirm that $\hat F_{(4)}$ takes the same
form as in (3.27) in \cite{Guarino:2015vca},
with the terms implicit in the ellipses
being   zero in our simplified setting.
This step can be checked
making use of the relations
(2.7), (2.8), (2.9) in \cite{Guarino:2015qaa}.
The next task  
is then to evaluate the quantities
$\cH_{(4)}^{IJ}$, $\cH_{(3)I}{}^J$,
$\tilde \cH_{(2)IJ}$ in 
(3.27) of \cite{Guarino:2015vca}.
This can be done using
(2.23), (2.21), (2.19)
in \cite{Guarino:2015qaa}.
The latter can be unpacked
using (C.3), (C.9) in the same
reference, keeping in mind that
for the dyonic gauging
at hand 
the correct values of $\theta_{AB}$,
$\xi^{AB}$ are given in 
(C.15) of \cite{Guarino:2015qaa}.
We obtain
\begin{align}
\cH_{(3)I}{}^J & = - T^{-1}_{IK} *DT^{KJ}
+ \frac 17 \delta_I{}^J T^{-1}_{KL} *DT^{LK}
\ ,  \label{dyonic_H3} \\
\tilde \cH_{(2)IJ} & = - Y^{- \frac 27} 
T^{-1}_{IK} T^{-1}_{JL} * \cH_{(2)}^{KL}
\ , \\
\cH_{(4)}^{IJ} & = Y^{\frac 27} \Big[  
T^{IJ} {\rm Tr}\,T - 2 (TT)^{IJ}
\Big] \ .
\label{dyonic_H4}
\end{align}
Here $\cH_{(2)}^{KL}$ is the notation
of \cite{Guarino:2015qaa} for the field strengths
$F^{KL}$.
Interestingly, all $Y$ and $dY$
factors drop away
from $\cH_{(3)I}{}^J$.

Once $\hat F_{(4)}$ is computed,
we may turn to $e^{\frac 12 \hat \phi} \hat * \hat F_{(4)}$, which may be evaluated
with the help of the identities 
of section \ref{app_Hodge}.
Taking the Hodge star with respect to the metric
\eqref{dyonic_metric}
generates additional powers of $Y$.
These conspire with the powers of $Y$
in \eqref{dyonic_H3}-\eqref{dyonic_H4}
to ensure that $\hat * \hat F_{(4)}$
has a single overall power of $Y$,
which is precisely cancelled
by the prefactor
$e^{\frac 12 \hat \phi}$.
We thus confirm that
$e^{\frac 12 \hat \phi} \hat * \hat F_{(4)}$
does not contain $Y$.
Finally, we verify that 
$e^{\frac 12 \hat \phi} \hat * \hat F_{(4)}$
takes the form
specified in \eqref{dyonic_flux}, \eqref{dyonic_e6},
up to an overall constant normalization
factor.

\section{Scalar potential from reduction of the $D$-dimensional action}
\label{app_potential}

\subsection{Reduction of the action: cases without dilaton}
\label{app_no_dilaton}

\paragraph{Reduction of the Einstein-Hilbert term.}
We find it convenient to
write the metric ansatz 
\eqref{ansatz_with_T} in the form 
\be  
d\hat s^2_D = (yTy)^{b_1} d \bar s^2_D \ , \qquad
d\bar s^2_D =  ds^2_d
+ d\tilde s^2_n
\ , \qquad 
d\tilde s^2_n =  g^{-2} (yTy)^{b_2} \gamma_{mn} d\xi^m d\xi^n \ .
\ee  
Here $\xi^m$ are local coordinates on $S^n$, and we have introduced the notation
\be  
\gamma_{mn} = T^{-1}_{IJ} \partial_m y^I
\partial_n y^J \ .
\ee  
Geometrically, $\gamma_{mn}$
describes an ellipsoid, whose
principal axes are determined
by the eigenvalues of the matrix $T_{IJ}$.

Our first task is the evaluation
of the Ricci scalar of the metric
$d\hat s^2_D$. 
In a first step,
we express 
the Ricci scalar
of $d \hat s^2_D$ in terms of that
of $d \bar s^2_D$, making use
of the fact that these two
$D$-dimensional line elements
are related by a Weyl rescaling
by the factor $(yTy)^{b_1}$.
Let $\hat \mu$, $\hat \nu$
be curved indices in $D$ dimensions.
The following formula is useful,
\begin{align}   \label{Ricci_rescaling}
\text{if} \;\; \hat g_{\hat \mu \hat \nu} & = e^{2\varphi}
\bar g_{\hat \mu \hat \nu} \ , \nn \\ 
R[\hat g_{\hat \mu \hat \nu}] & = e^{-2\varphi} \, \Big[ 
R[\bar g_{\hat \mu \hat \nu}] - 
2   (D-1) \, \bar g^{\hat \mu \hat \nu} \, \overline \nabla_{\hat \mu} \overline \nabla_{\hat \nu} \varphi
- (D-1)(D-2) \, \bar g^{\hat \mu \hat \nu} \, \overline \nabla_{\hat \mu} \varphi \, \overline  \nabla_{\hat \nu} \varphi
\Big] \ . 
\end{align}
The symbol $\overline \nabla_{\hat \mu}$
denotes the Levi-Civita connection
of the $D$-dimensional metric
$d\bar s^2_D$.
We apply \eqref{Ricci_rescaling} with $e^{2\varphi} = (yTy)^{b_1}$.

Next, we observe that
the metric $d\bar s^2_D$
is a direct product between
the $d$-dimensional metric
$ds^2_d = g_{\mu\nu} dx^\mu dx^\nu$
and the $n$-dimensional metric  
$d\tilde s^2_n = \tilde g_{mn} d\xi^m d\xi^n$.
(We have introduced
$\mu$, $\nu$, which are
curved indices in $d$ dimensions.)
This holds because we are
working under the simplifying
assumptions \eqref{simpler_case}
of no gauge fields and constant $T$.
As a result, we have the simple formula
\be  
R[\bar g_{\hat \mu \hat \nu}]
= R[g_{\mu\nu}] + R[\tilde g_{mn} ] \ .
\ee   
Furthermore, the $D$-dimensional
Levi-Civita connection 
$\overline \nabla_{\hat \mu}$
also splits in a trivial way,
\be  
\overline \nabla_{\hat \mu} = (
\overline \nabla_{ \mu} , 
\overline \nabla_{m}
)=
 (
 \nabla_{ \mu} , 
\tilde  \nabla_{m}
)  \ .
\ee  
Here  
$\nabla_{ \mu}$ is the Levi-Civita
connection of $g_{\mu\nu}$
and $\tilde \nabla_m$
is that of $\tilde g_{mn}$.
The conformal factor $e^{2\varphi} = (yTy)^{b_1}$ in \eqref{Ricci_rescaling}
depends on the $S^n$ coordinates
only (because we are assuming $T$ is constant). As a result, we can write
\begin{align}
R[\hat g_{\hat \mu \hat \nu}] & = e^{-2\varphi} \, \Big[ 
R[g_{\mu\nu}]
+ R[\tilde g_{mn}]
- 
2   (D-1) \, \tilde g^{m n} \, \tilde \nabla_{m} \tilde \nabla_{n} \varphi
- (D-1)(D-2) \, \tilde g^{m n} \, \tilde \nabla_{m} \varphi \, \tilde  \nabla_{n} \varphi
\Big] \ . 
\end{align}

We may now proceed 
by writing quantities associated to
$\tilde g_{mn}$ in terms of
quantities associated to $\gamma_{mn}$.
The two metrics are related
as $\tilde g_{mn} = e^{2 \varphi'} \gamma_{mn}$ with 
$e^{2 \varphi'} = g^{-2} (yTy)^{b_2}$.
For the Ricci scalar
$R[\tilde g_{mn}]$ we can use
a formula completely analogous to 
\eqref{Ricci_rescaling}.
We also have immediately
$\tilde \nabla_m \varphi = \partial_m \varphi$.
The Laplacian term
$\tilde g^{m n} \, \tilde \nabla_{m} \tilde \nabla_{n} \varphi$ can be
addressed with the help of the identity
\begin{align}   \label{Laplacian_rescaling}
\text{if} \;\; \tilde g_{m n}   = e^{2\varphi'}
 \gamma_{m n} \ , \;
\tilde g^{m n} \, \tilde \nabla_{m} \tilde \nabla_{n} \varphi   = e^{-2\varphi'} \, \Big[  
\gamma^{mn} \nabla^{(\gamma)}_m \nabla^{(\gamma)}_n \varphi
+ (n-2) \gamma^{mn} \partial_m \varphi 
\partial_n \varphi ' 
\Big] \ . 
\end{align}
We have introduced the notation
$\nabla^{(\gamma)}_m$
for the Levi-Civita connection
of $\gamma_{mn}$.

Retracing all the steps outlined
above,
we arrive at the following formula
for the Ricci scalar of the 
$D$-dimensional metric $d\hat s^2_D$,
\begin{align} \label{Rhat_eq}
R[\hat g_{\hat \mu \hat \nu}] & = (yTy)^{-b_1} \, R[g_{\mu\nu}] + g^2 \, (yTy)^{-b_1-b_2} \, \bigg[  R[\gamma_{mn}]  
+ \cK_{\rm (L)} \, \gamma^{mn} \,  \nabla^{(\gamma)}_m  \nabla^{(\gamma)}_n \log(yTy)   \nn  \\
& \qquad + \cK_{\rm (G)} \, \gamma^{mn} \, \partial_m \log(yTy)
\partial_n  \log(yTy) \bigg] \ .
\end{align}
We have introduced the constants
\begin{align}
 \cK_{\rm (L)} & = b_1 - b_1 D + b_2 - b_2 n \  , \nn \\
  \cK_{\rm (G)} & = - \frac{b_1^2}{4} (D-1)(D-2)
- \frac{b_2^2}{4} (n-1)(n-2)
- \frac{b_1 b_2}{2} \, (D-1)(n-2) \ .
\end{align}
The volume form of the $D$-dimensional
metric $d \hat s^2_D$ is written 
in terms of the $d$-dimensional metric
$g_{\mu\nu}$ and the $n$-dimensional
metric $\gamma_{mn}$ as
\be \label{volhat_eq}
\sqrt{- \hat g_{ D}} = g^{-n}\, (yTy)^{b_1 D/2 + b_2 n/2 } \, \sqrt{- g_{d}} \, \sqrt{ \gamma} \ .
\ee 
The $D$-dimensional
Einstein-Hilbert Lagrangian
$\sqrt{- \hat g_{ D}} R[\hat g_{\hat \mu \hat \nu}]$ is readily written
by combining 
\eqref{Rhat_eq} and \eqref{volhat_eq}.

We are free to add to 
$\sqrt{- \hat g_{ D}} R[\hat g_{\hat \mu \hat \nu}]$ any total divergence on $S^n$,
of the form
$\sqrt {\gamma} \nabla^{(\gamma)}_m (\dots)^m$,
because we are interested in the
integral of the $D$-dimensional
action over $S^n$.
If we add a term of the form
\be  
\sqrt {\gamma} \nabla^{(\gamma)}_m 
\Big( 
\sqrt{- g_d}  \gamma^{mn} \nabla^{(\gamma)}_n
(yTy)^{\frac{b_1 D}{2}-b_1+\frac{b_2 n}{2}-b_2}
\Big) 
\ee  
with the appropriate constant prefactor,
we   eliminate the
Laplacian term
$\gamma^{mn} \,  \nabla^{(\gamma)}_m  \nabla^{(\gamma)}_n \log(yTy)$.
We thus get the simpler expression
\begin{align}
& \sqrt{- \hat g_{ D}} R[\hat g_{\hat \mu \hat \nu}]
 = 
  g^{-n}(yTy)^{-b_1
  + b_1 D/2 + b_2 n/2} \sqrt{- g_{d}} \, \sqrt{ \gamma} R[g_{\mu\nu}]
    \\
  & +g^{-n+2} \, (yTy)^{-b_1-b_2
  + b_1 D/2 + b_2 n/2} \sqrt{- g_{d}} \, \sqrt{ \gamma} \bigg[  R[\gamma_{mn}] 
  + \cK'_{\rm (G)} \, \gamma^{mn} \, \partial_m \log(yTy)
\partial_n  \log(yTy) \bigg] \ , \nn
\end{align}
where the new constant $\cK'_{\rm (G)}$
is given by
\be  
\cK'_{\rm (G)} = 
\cK_{\rm (G)} - \cK_{\rm (L)} \left(
\frac{b_1 D}{2}-b_1+\frac{b_2 n}{2}-b_2
\right) \ . 
\ee  

In a final step, we rewrite the quantities
expressed in terms of the ellipsoid metric
$\gamma_{mn}$ in terms of the round metric
$\accentset{\circ}{g}_{mn}$ defined in \eqref{round_metric_def}.
The following identities are useful,
\begin{align}
\sqrt{\gamma}   = (yTy)^{1/2} \, \sqrt{
\accentset{\circ}{g}
} \ , \qquad 
 \gamma^{mn} \, \partial_m (yTy) \partial_n  (yTy)    = 
  4 \, (yT^3y) - 4 \, \frac{(yT^2y)^2}{yTy} \ .
\end{align}
We also need the expression of the
Ricci scalar of the ellipsoid metric
\cite{Nastase:2000tu},
\be \label{ellipsoid_Ricci}
R[\gamma_{mn}] = \frac{( {\rm Tr} \, T )^2}{ yTy} - \frac{{\rm Tr} \, T^2}{yTy}
- \frac{2 \, ({\rm Tr} \, T) \, (yT^2y)}{(yTy)^2} + \frac{2 \,  yT^3y }{(yTy)^2} \ .
\ee
Combining all the above ingredients,
we finally arrive at the result
quoted in 
\eqref{EH_reduction}, \eqref{cG_expression}.

As explained in the main text,
the values of the parameters $b_1$,
$b_2$ are fixed according to 
\eqref{fix_b1_b2}.
After that, we find it convenient to
add another total divergence on $S^n$.
More precisely, we add a constant multiple of
\begin{align}
& \sqrt {\gamma} \nabla^{(\gamma)}_m 
\Big( 
\sqrt{- g_d}  \gamma^{mn} \nabla^{(\gamma)}_n
(yTy)^{1/2}
\Big) 
 = \nn \\
& = 2 \sqrt{- g_d}\sqrt{ \accentset{\circ}{g} } \bigg[  
\frac{(yT^2y)^2 }{(yTy)^2} - \frac{(yT^3y)}{yTy}   + \frac 12 \, {\rm Tr}(T^2)
- \frac { ( {\rm Tr} \, T ) \, (yT^2 y) }{2 \, (yTy)} 
\bigg]   \ .
\end{align} 
Adding this total divergence allows us
to cancel the terms
with $yT^3y$
and complete a perfect square,
thus arriving
at the result 
\eqref{EH_term} quoted in the main text.

\paragraph{Reduction of kinetic term
for the flux.}

The kinetic term for $\hat \cF_n$
is proportional to the quantity
\be   \label{flux_squared}
\sqrt{ -\hat g_D} | \hat \cF_n  |^2  
=\frac{1}{n!}\sqrt{ -\hat g_D} \hat \cF_{\hat \mu_1 \dots \hat \mu_n}
\hat \cF_{\hat \nu_1 \dots \hat \nu_n}
\hat g^{\hat \mu_1 \hat \nu_1}
\dots 
\hat g^{\hat \mu_n \hat \nu_n} \ .
\ee 
The ansatz \eqref{f_definition}
for $\hat \cF_n$ implies
that the only non-zero components of
$\hat \cF_n$ are
\be    \label{flux_components}
\hat \cF_{p_1 \dots p_n} = f(y,T)
\accentset{\circ}{\epsilon}_{p_1 \dots p_n} ,
\ee  
where $\accentset{\circ}{\epsilon}$
is the volume form of the round metric
$\accentset{\circ}{g}_{mn}$
on $S^n$.
In a first step, we can rewrite
\eqref{flux_squared} in terms of the $D$-dimensional
metric $d \bar s^2_D$, generating
an overall $yTy$ prefactor,
\be  
\sqrt{ -\hat g_D} | \hat \cF_n  |^2  
=\frac{1}{n!}(yTy)^{b_1 D/2 - b_1 n}\sqrt{ -\hat g_D} \hat \cF_{\hat \mu_1 \dots \hat \mu_n}
\hat \cF_{\hat \nu_1 \dots \hat \nu_n}
\hat g^{\hat \mu_1 \hat \nu_1}
\dots 
\hat g^{\hat \mu_n \hat \nu_n} \ .
\ee 
Next, we use the fact that
$d\bar s^2_D$ is a direct product metric,
$d\bar s^2_D = ds^2_d + d\tilde s^2_n$,
and that $\hat \cF_n$ has only internal legs,
\begin{align}  
\sqrt{ -\hat g_D} | \hat \cF_n  |^2  
=\frac{1}{n!}(yTy)^{b_1 D/2 - b_1 n}
\sqrt{ -  g_d} \sqrt{\tilde g} \hat \cF_{p_1 \dots p_n}
\hat \cF_{q_1 \dots q_n}
\tilde g^{p_1 q_1}
\dots 
\tilde g^{p_n q_n} \ .
\end{align}
We proceed by trading 
the metric $\tilde g_{mn}$
with the ellipsoid metric $\gamma_{mn}$
and using \eqref{flux_components},
\begin{align}  
\sqrt{ -\hat g_D} | \hat \cF_n  |^2  
=\frac{1}{n!} g^n (yTy)^{b_1 D/2 - b_1 n
- b_2 n /2  }
\sqrt{ -  g_d} \sqrt{\gamma}
f^2  \accentset{\circ}{\epsilon}_{p_1 \dots p_n}
\accentset{\circ}{\epsilon}_{q_1 \dots q_n}
\gamma^{p_1 q_1}
\dots 
\gamma^{p_n q_n} \ .
\end{align}
To compute the contraction of two
\emph{round} volume forms 
$\accentset{\circ}{\epsilon}$
with the \emph{ellipsoid} inverse metric,
we use $\det \gamma_{mn} = (yTy) \det 
\accentset{\circ}{g}_{mn}$
(valid for unimodular $T$)
to write
\be  
\accentset{\circ}{\epsilon}_{p_1 \dots p_n}
\accentset{\circ}{\epsilon}_{q_1 \dots q_n}
\gamma^{p_1 q_1}
\dots 
\gamma^{p_n q_n}
= (yTy)^{-1} {\epsilon}^{(\gamma)}_{p_1 \dots p_n}
{\epsilon}^{(\gamma)}_{q_1 \dots q_n}
\gamma^{p_1 q_1}
\dots 
\gamma^{p_n q_n}
= n! (yTy)^{-1} \ ,
\ee  
where ${\epsilon}^{(\gamma)}$ is the volume
form of the ellipsoid metric
$\gamma_{mn}$.
We thus reproduce the result
\eqref{flux_reduction_res} quoted in the main text.

\subsection{Integrals on $S^n$ and independence on $T$}
\label{app_integrals_and_T}

Our goal is to determine under which conditions
the following quantity
has an integral over $S^n$ that is independent of $T$
(provided $\det T = 1$),
\be   \label{h_function}
h(T,y) =
(yTy)^{-k/2 + u  }\bigg[
\frac{ yT^2y  }{(yTy)^2} - \frac{4-k}{8} \frac{{\rm Tr} \, T }{yTy}
\bigg]  \ .
\ee  
We have included
two independent constant parameters $k$, $u$,
in order to be able to apply the present argument
to the case with a dilaton.

We consider a $T$ matrix of the form $T_{IJ}= \exp(t)_{IJ}$,
with ${\rm Tr}\,t = 0$, so that $\det T  =1$.
We can expand $h(e^t,y)$ in a power series around $t=0$.
The integral over $S^n$ can be performed
order by order with the help
of identities 
\be     \label{integration_identities}
\int_{S^n} d^n \xi \sqrt{ \accentset{\circ}{g}}
y^{I_1} \dots y^{I_{2p}}
= \frac{(2p-1)!!  \, \delta^{( I_1 I_2} \delta^{I_3 I_4} \dots \delta^{I_{2p-1}
I_{2p} )
} }{ (n+1) (n+3) \dots (n+2p-1)}
\ , \qquad 
p=1,2,3,\dots 
\ee


For simplicity, let us first freeze $u=0$,
keeping $k$ as only free parameter.
This is enough for applications to the cases without
dilaton.
We find 
\begin{align}
& \frac{1}{\cV_n} \int_{S^n} d^n \xi  \sqrt{  \accentset{\circ}{g} }   h(e^t,y)   = 
\frac{1}{8} ((k-4) n+k+4)
 \\
& + \frac{k   \left(k^2 (n+1)-k n (n+2)+7 k+4 (n-3) (n+1)\right)}{32 (n+1) (n+3)}
{\rm Tr}(t^2)
\nn \\
& -\frac{k   (k-n-1) \left(2 k^2 (n+1)+k (19-(n-2) n)+4 (n-7) (n+1)\right)}{96
   (n+1) (n+3) (n+5)}
{\rm Tr}  (t^3) + \cO(t^4) \ . \nn
\end{align}
We see that the only way to ensure that the
result is independent of $t$, without
imposing any restriction on $t$
(except tracelessness), is to select
\be  
k = 0 \ .
\ee  
We know that \eqref{h_function}
with $(k,u) = (0,0)$
is related to a total derivative,
thanks to the identity
\eqref{why_it_is_sufficient}, repeated here for convenience,
\be   
 \epsilon y (dy)^n + d\bigg[
\frac{n}{n-1} \frac{1}{yTy} \epsilon y (Ty) (dy)^{n-1}
\bigg] = - \frac{2}{n-1}\bigg[ \frac{ yT^2y  }{(yTy)^2} - \frac 12 \frac{ {\rm Tr} \, T  }{yTy}
\bigg]  \epsilon y (dy)^n  \ .
\ee  
This relation demonstrates that
\be   \label{first_case_res}
(k,u) = (0,0) \quad \Rightarrow \quad \int_{S^n}
d^n \xi \sqrt{ \accentset{\circ}{g}   }  h(y,T) = - \frac{n-1}{2} \cV_n \ ,
\ee  
which is indeed non-zero for $n \ge 2$
and independent of $T$.

Let us now repeat the analysis keeping both $k$
and $u$ as independent parameters,
as required for applications to the cases with a dilaton.
We have studied the integral of \eqref{h_function}
up to fourth order in $t$.
By collecting the coefficients
of the ${\rm Tr} (t^2)$, ${\rm Tr} (t^3)$, ${\rm Tr} (t^4)$,
and $[{\rm Tr} (t^2)]^2$ structures,
and setting them to zero,
we obtain a set of relations among $k$, $u$, $n$,
which admit the following solutions,
\begin{align}   \label{three_options}
(k,u) = (0,0) \ ,  \quad 
(k,u)  ={ \scriptstyle ( \frac{4 (n-1)}{n+1} , 
-\frac{(n-3) (n-1)}{2 (n+1)} )
}\ ,    \quad
(k,u)  = { \scriptstyle ( 
\frac{4 (n+1)}{n+3} , -\frac{(n-1) (n+1)}{2 (n+3)}
) }  \ .
\end{align}
These three options are the only possible 
values of $(k,u)$ for which the integral
of \eqref{h_function} has a chance of being
independent of $T$, for unimodular $T$.
We now study each case and prove that the integral
has indeed this property.

We have already encountered the case $(k,u) = 0$,
and we have already demonstrated in \eqref{first_case_res}
that for these values of $k$, $u$ the integral
of $h$ is indeed independent of $T$.

For the case $(k,u) = { \scriptstyle ( \frac{4 (n-1)}{n+1} , 
-\frac{(n-3) (n-1)}{2 (n+1)} )
}$ we   use   the relation
\be  
 d\bigg[
(yTy)^{- \frac{n+1}{2}} \epsilon y (Ty) (dy)^{n-1}
\bigg] = - \frac{n+1}{n} (yTy)^{1- \frac{n+1}{2}} \bigg[ 
\frac{yT^2 y}{(yTy)^2}
- \frac{1}{n+1} \frac{{\rm Tr} \, T}{yTy} 
\bigg] \epsilon y (dy)^n \ .
\ee  
We conclude that
\be  
(k,u) = { \scriptstyle ( \frac{4 (n-1)}{n+1} , 
-\frac{(n-3) (n-1)}{2 (n+1)} )
} \quad \Rightarrow \quad \int_{S^n}d^n \xi  \sqrt{ \accentset{\circ}{g}   }  h(y,T) \equiv 0 \ .
\ee 
While indeed independent of $T$,
a flux that is identically zero is not 
acceptable, because it does not reproduce
the non-zero flux of the case $T = \mathbb I$.

The third option in \eqref{three_options} requires
a more careful analysis.
We start from
\begin{align}  
\sqrt{ \accentset{\circ}{g}  } \accentset{\circ}{\nabla}^m 
\accentset{\circ}{\nabla}_m (yTy)^\lambda & = 
\sqrt{ \overline g }\, \Big[ 
2 \, \lambda \, (yTy)^{\lambda -1} \, {\rm Tr} \, T
- 2 \, \lambda \, (n+1) \, (yTy)^\lambda
\nn \\
& + 4 \, \lambda \, (\lambda -1) \, (yTy)^{\lambda -2} \, (yT^2y)
- 4 \, \lambda \, (\lambda -1) \, (yTy)^\lambda \Big] \ ,
\end{align}  
where $\accentset{\circ}{\nabla}^m 
\accentset{\circ}{\nabla}_m$
 is the scalar Laplace operator
constructed with the round metric on $S^n$
and $\lambda$ is a real constant.
If we specialize to $\lambda = -(n+1)/2$,
we obtain
\be  
\sqrt{ \accentset{\circ}{g}  } \accentset{\circ}{\nabla}^m 
\accentset{\circ}{\nabla}_m (yTy)^\lambda
= (n+1)(n+3) \sqrt{ \accentset{\circ}{g}  } (yTy)^{ - \frac{n+1}{2}} \bigg[ 
\frac{yT^2y}{(yTy)^2}
- \frac{1}{n+3} \frac{ {\rm Tr} \, T }{yTy} 
- \frac{2}{n+3}
\bigg]    \ .
\ee  
Since the LHS is a total divergence, we conclude that
\be \label{third_res_temp}
(k,u) = { \scriptstyle ( 
\frac{4 (n+1)}{n+3} , -\frac{(n-1) (n+1)}{2 (n+3)}
) } \quad 
\Rightarrow 
\quad
\int_{S^n} \sqrt{  \accentset{\circ}{g} }  h(y,T)
= \frac{2}{n+3} \int_{S^n}d^n \xi  \sqrt{  \accentset{\circ}{g} }
(yTy)^{- \frac{n+1}{2}} \ .
\ee  
The integral on the RHS can be evaluated as follows.
We start from the standard Gaussian integral
\be   \label{first_way}
\int_{\mathbb R^{n+1}} d^{n+1} x e^{ - \frac 12 x T x }
= (2\pi)^{\frac{n+1}{2}} (\det T)^{- \frac 12} \ ,
\ee  
where $x^I$ are Cartesian coordinates on $\mathbb R^{n+1}$.
We introduce polar coordinates
$x^I = r y^I$, with $y^I y_I =1$.
We can then write
\begin{align}  
\int_{\mathbb R^{n+1}} d^{n+1} x e^{ - \frac 12 x T x }
& =   \int_{ S^{n}} d^n \xi
\sqrt{ \accentset{\circ}{g} } 
\int_0^\infty dr  r^n  e^{ - \frac 12 r^2 (y T y) } 
\nn \\
& =  \int_{ S^{n}}  d^n \xi
\sqrt{ \accentset{\circ}{g} }  2^{\frac{n-1}{2}} (yTy)^{- \frac{n+1}{2}} \Gamma(
\tfrac{n+1}{2}
)
\ . \label{second_way}
\end{align}
Comparing \eqref{first_way} and \eqref{second_way}
and recalling \eqref{unit_sphere}, 
we find
\be  
 \int_{ S^{n}}  d^n \xi
\sqrt{ \accentset{\circ}{g} } 
(yTy)^{- \frac{n+1}{2}}  = \cV_n (\det T)^{- \frac 1 2} \ .
\ee  
Using \eqref{third_res_temp} we conclude
\be \label{third_res}
(k,u) = { \scriptstyle ( 
\frac{4 (n+1)}{n+3} , -\frac{(n-1) (n+1)}{2 (n+3)}
) } \quad 
\Rightarrow 
\quad
\int_{S^n} \sqrt{  \accentset{\circ}{g} }  h(y,T)
= \frac{2}{n+3} \cV_n (\det T)^{- \frac 1 2} \ .
\ee  
The result is not identically zero and  depends on $T$.
The dependence on $T$, however,
drops out after imposing the
unimodularity constraint.

To summarize: the integral of $h$ in \eqref{h_function}
is independent of $T$ for unimodular $T$ if and only
if one of the following three cases is realized:
\begin{itemize}
\item $(k,u) = (0,0)$: the integral of $h$ is non-zero
and independent of $T$ without using the condition 
$\det T = 1$;
\item $(k,u) = { \scriptstyle ( \frac{4 (n-1)}{n+1} , 
-\frac{(n-3) (n-1)}{2 (n+1)} )
}$: the integral of $h$ is identically zero for
any symmetric $T$ without using the condition 
$\det T = 1$;
\item $(k,u) = { \scriptstyle ( 
\frac{4 (n+1)}{n+3} , -\frac{(n-1) (n+1)}{2 (n+3)}
) }$: the integral of $T$ is a non-zero constant
times $(\det T)^{- \frac 12}$, and thus 
independent of $T$ using $\det T = 1$.
\end{itemize}

\subsection{Features of non-trivial integrals over $S^n$}
\label{app_integral_example}

The reduction of the Einstein-Hilbert 
term yields the integrand $\cI$
in \eqref{EH_term}. The terms with a non-trivial
$y$ dependence are collected in the perfect square
$[ \frac{ yT^2y }{ yTy } 
- \frac {4 - k}{8}   {\rm Tr} \, T 
 ]^2$. 
This is a rational function of $y$, $T$,
homogeneous of degree 2 under
a formal rescaling $T_{IJ} \rightarrow \lambda T_{IJ}$. (In the present discussion,
we relax the unimodularity constraint on $T$
and study integrals of expressions
such as $\frac{ yT^2y }{ yTy }$
as functions of an arbitrary symmetric $T$.)  If we integrate
$[ \frac{ yT^2y }{ yTy } 
- \frac {4 - k}{8}   {\rm Tr} \, T 
 ]^2$
 over $S^n$ with measure $\sqrt{ \accentset{\circ}{g} }$, 
 we get a 
 a function of $T_{IJ}$ which is manifestly
homogeneous of degree 2. It is not, however, a quadratic
function, nor a rational function
of the entries of $T$.

This can be verified explicitly
in the example $n=2$,
choosing for simplicity
a diagonal $T$ of the form
\be  
T_{IJ} = {\rm diag}( t, t, s ) \ ,
\ee 
where $t$, $s$ are positive parameters.
The constrained coordinates $y^I$ may be parametrized
in terms of an interval coordinate $-1 \le \mu \le 1$
and an angle $\phi$ with period $2\pi$,
\be 
y^1 = \sqrt{1- \mu^2} \cos \phi \ , \qquad
y^2 = \sqrt{1- \mu^2} \sin \phi \ , \qquad
y^3 = \mu \ .
\ee  
The round metric on $S^2$ in these coordinates
reads $d\accentset{\circ}{s}^2 = \frac{d\mu^2}{1-\mu^2}
+ (1-\mu^2) d\phi^2$ and therefore
$\sqrt{  \accentset{\circ}{g} }=1$.
We also check that the quantity
$[ \frac{ yT^2y }{ yTy } 
- \frac {4 - k}{8}   {\rm Tr} \, T 
]^2$ is independent of $\phi$,
due to the fact that the first two diagonal
entries of $T$ are equal.
We are thus left to compute
an integration in the variable $\mu$ only.
Considering for
definiteness the case $s>t$,
we find\footnote{Our findings in this explicit
example are not compatible with some
of the identities quoted in \cite{Nastase:2000tu}.}
\be   \label{integral_example}
\frac{1}{2\pi} \int_{S^2} \sqrt{ 
\accentset{\circ}{g}
}  
\bigg[ \frac{ yT^2y }{ yTy } 
- \frac {4 - k}{8}   {\rm Tr} \, T 
 \bigg]^2
= s t + \frac{((k+4) s+2 k t)^2}{32} -\frac{s  ((k+2) s+2 k t) \arctan
   \sqrt{\frac st -1}}{2 \sqrt{\frac st -1}} \ .
\ee   
The RHS is homogeneous of degree 2
under a simultaneous rescaling
$(t,s) \rightarrow (\lambda t, \lambda s)$,
but it is not a rational function of $t$, $s$.

\bibliographystyle{JHEP}
\bibliography{refs}

\end{document}